\documentclass[12pt,a4paper]{article}

\usepackage[latin1]{inputenc}
\usepackage{amsmath}
\usepackage{graphicx}
\usepackage{verbatim}
\usepackage{multirow}
\usepackage[labelfont=bf,textfont={sl,bf}]{subfig}

\newcommand{\bc}{\begin{center}}
\newcommand{\ec}{\end{center}}
\newcommand{\be}{\begin{equation}}
\newcommand{\ee}{\end{equation}}
\newcommand{\bea}{\begin{eqnarray}}
\newcommand{\eea}{\end{eqnarray}}
\newcommand{\beqn}{\begin{eqnarray}}
\newcommand{\eeqn}{\end{eqnarray}}
\newcommand{\ba}{\begin{array}}
\newcommand{\ea}{\end{array}}
\newcommand{\bal}{\begin{aligned}}
\newcommand{\eal}{\end{aligned}}
\newcommand{\ben}{\begin{enumerate}}
\newcommand{\een}{\end{enumerate}}
\newcommand{\bitem}{\begin{itemize}}
\newcommand{\eitem}{\end{itemize}}

\newcommand{\pa}{\partial}
\newcommand{\fr}{\frac}
\newcommand{\bpmatrix}{\begin{pmatrix}}
\newcommand{\epmatrix}{\end{pmatrix}}

\newcommand{\crn}{\nonumber \\}

\newcommand{\al}{\alpha}

\newcommand{\ga}{\gamma}

\newcommand{\de}{\delta}
\newcommand{\De}{\Delta}
\newcommand{\eps}{\epsilon}

\newcommand{\si}{\sigma}
\newcommand{\Si}{\Sigma}

\newcommand{\DR}{{\text{DR}}}
\newcommand{\MR}{{\text{MR}}}

\newcommand{\MSb}{{\overline{\text{MS}}}}
\newcommand{\DIS}{{\text{DIS}}}

\newcommand{\dd}{{\rm{d}}}
\newcommand{\hsigma}{{\hat{\sigma}}}

\newcommand{\eq}[1]{Eq.~(\ref{#1})}
\newcommand{\bib}[1]{Ref.~\cite{#1}}
\newcommand{\fig}[1]{Fig.~\ref{#1}}
\newcommand{\tab}[1]{Table~\ref{#1}}
\newcommand{\sect}[1]{Section~\ref{#1}}

\newcommand{\braket}[1]{\left(#1\right)}
\newcommand{\gev}{{\unskip\,\text{GeV}}}
\newcommand{\tev}{{\unskip\,\text{TeV}}}

\newcommand{\qqwwz}{q\bar{q}\to W^+W^- Z}

\newcommand{\yywwz}{\gamma\gamma\to W^+W^- Z}

\newcommand{\hs}{\hspace*{3mm}}
\newcommand{\ie}{{\it i.e.}}
\newcommand{\eg}{{\it e.g.}}

\def\slashepi{\epsilon_i\kern -.720em {/}}
\def\slashpi{p_i\kern -.600em {/}}

\begin{document}

\begin{titlepage}
\vspace*{0.1cm}
\rightline{KA-TP-19-2013}
\rightline{MPP-2013-220}
\rightline{SFB/CPP-13-52}

\vspace{1mm}
\begin{center}

{\Large{\bf NLO corrections to $WWZ$ production at the LHC}}

\vspace{.5cm}

DAO Thi Nhung$^{a,b}$, LE Duc Ninh$^{a,b}$ and Marcus M. WEBER$^c$

\vspace{4mm}
{\it $^a$Institut f\"ur Theoretische Physik, Karlsruher Institut f\"ur  Technologie, \\
D-76128 Karlsruhe, Germany\\
$^b$ Institute of Physics, Vietnam Academy of Science and Technology, \\
10 Dao Tan, Ba Dinh, Hanoi, Vietnam}

{\it $^c$Max-Planck-Institut f\"ur Physik (Werner-Heisenberg-Institut), \\
D-80805 M\"unchen, Germany}

\vspace{10mm}
\abstract{The production of $W^+W^-Z$ at the LHC is an important process to test the quartic gauge couplings
of the Standard Model as well as an important background for new physics searches. A good theoretical
understanding at next-to-leading order (NLO) is therefore valuable. In this paper, we present the calculation of
the NLO electroweak (EW) correction to this channel with on-shell gauge bosons in the final state.
It is then combined with the NLO QCD correction to get the most up-to-date prediction.
We study the impact of these corrections on the total cross section and some distributions.
The NLO EW correction is small for the total
cross section but becomes important in the high energy regime for
the gauge boson transverse momentum distributions.}

\end{center}
%\vspace*{\fill} {\bf \Large \today}

\normalsize

\end{titlepage}

\section{Introduction}
\label{intro}
The program to check the Standard Model (SM) is on good course with
the recent discovery of a new boson with a mass of about $125\gev$ at
the ATLAS \cite{:2012gk} and CMS \cite{:2012gu} experiments.
The present data seem to indicate that this new particle is consistent with
the long-sought SM Higgs boson, whose existence is
a prediction of the SM. Once the particle list is confirmed and their masses are
measured, we have to make sure that all the SM couplings are consistent with the data.
In this project, we have to check the quartic couplings of gauge bosons, which are
renormalizable and occur in the SM Lagrangian as a consequence of non-Abelian gauge
symmetry.

If we take a proton-proton collider and ask the question what is a good process to test
the quartic couplings $W^+W^-ZZ$ and $W^+W^-Z\gamma$ then we find that there are two mechanisms at tree level.
The four-point vertex is attached to either one quark line or two quark lines. 
In this paper we consider the former with three massive gauge bosons in the
final state, namely the process $pp\rightarrow W^+W^-Z$. In addition, this process 
is an important background for new physics searches.

The tree-level requirement is important to have high sensitivity to the couplings. In order to
compare the SM prediction with experimental data, the tree-level calculation is, however, not
good enough since it suffers from large theoretical uncertainties. A full next-to-leading-order (NLO)
calculation including both QCD and EW corrections is needed to reduce the uncertainty and
to understand the quantum-loop effects. The NLO QCD corrections have been
calculated by two groups: in \bib{Hankele:2007sb}
including leptonic decays of the gauge bosons and in \bib{Binoth:2008kt} in the heavy Higgs limit. In this paper
we recalculate the QCD corrections and, for the first time, the full NLO EW corrections to the on-shell $W^+W^-Z$ production
at the large hadron collider (LHC) are calculated.

The paper is organized as follows. The calculation of NLO QCD and EW corrections is discussed
in \sect{cal_details}. The definition of hadronic cross section is also given there. In
\sect{sect_result}, numerical results for the total cross section and some representative distributions
are presented. We discuss also the use of jet veto to reduce large QCD correction. Conclusions are
found in the last section. In the appendix we provide results at the amplitude squared level for 
a random phase-space point to facilitate comparisons with our results. 

%%%%%%%%%%%%%%%%%%%%%%%%%%%%%%%%%%%%%%%%%%%%
\section{Calculational details}            %
\label{cal_details}                        %
%%%%%%%%%%%%%%%%%%%%%%%%%%%%%%%%%%%%%%%%%%%%
\begin{figure}[h]
  \centering
 \subfloat[]{ \includegraphics[width=0.8\textwidth,height=0.15\textwidth]
{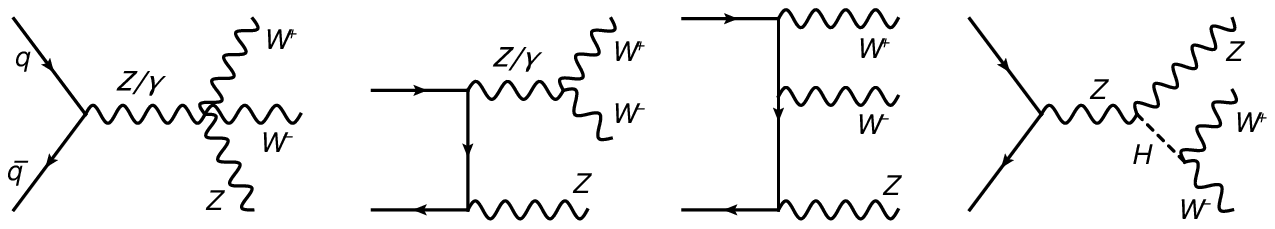}\label{qqWWZ_tree_diag}}\\
 \subfloat[]{ \includegraphics[width=0.8\textwidth,height=0.15\textwidth]
{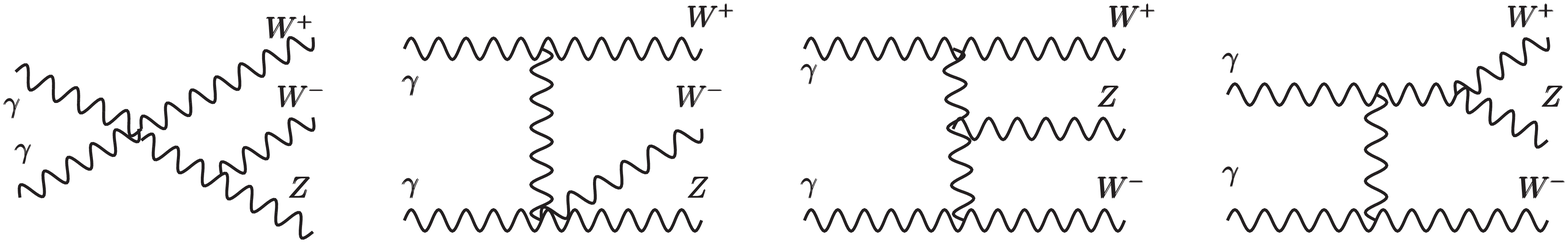} \label{yyWWZ_tree_diag}}\\
  \caption{ Representative tree-level diagrams for the
$\qqwwz$ subprocesses (a) and the $\yywwz$ subprocess (b).}
\end{figure}
The tree-level subprocesses are
\bea
\bar{q} + q & \to & W^+ + W^- + Z, \label{qqwwz} \\
\bar{b} + b & \to & W^+ + W^- + Z, \label{bbwwz} \\
\gamma + \gamma & \to & W^+ + W^- + Z, \label{yywwz}
\eea
where $q$ stands for the light quarks ($u,d,c,s$) if not otherwise stated.
The $q\bar{q}$ contributions are dominant and their Feynman diagrams
can be divided into four distinct topologies as depicted in \fig{qqWWZ_tree_diag}.
It should be noted that the $s$-channel diagrams with an intermediate Higgs boson
are included in our calculation.
They alone form a gauge invariant set.
The Higgs contribution including interference effects is less than $1\%$ at leading order (LO) for
$M_H = 125\gev$. Since the bottom-quark and photon distribution
functions are much smaller than those of the light quarks,
the $b\bar{b}$ and $\gamma\gamma$ contributions are much less important.
We therefore include them only at LO.
In \fig{yyWWZ_tree_diag}, a representative set of tree-level diagrams for $\yywwz$ is presented.

In the following we discuss the NLO QCD and
EW corrections to the subprocesses~(\ref{qqwwz}).
We will define the various sub-corrections at NLO, namely the QCD virtual,
gluon-radiated and gluon-induced corrections for the QCD case and the
EW virtual, photon-radiated and photon-induced corrections for the EW
case. These sub-corrections are ultraviolet (UV) and infrared (IR)
finite, but are dependent on the regularization scheme. The final
results, {\it i.e.} the sum of those sub-corrections, are
regularization-scheme independent. The separation will provide more
insights into the QCD and EW corrections.
%%%%%%%%%%%%%%%%%%%%%%%%%%%%%%%%%%
\subsection{NLO QCD corrections} %
\label{cal_qcd}                  %
%%%%%%%%%%%%%%%%%%%%%%%%%%%%%%%%%%
The NLO QCD contribution contains the virtual and real-emission corrections.
The virtual Feynman diagrams with an extra gluon in the loops include pentagon diagrams up to
rank four. The one-loop tensor
integrals are calculated using Passarino-Veltman reduction~\cite{Passarino:1978jh}
for up to four-point diagrams and the method of \bib{Denner:2005nn} (see also \bib{Binoth:2005ff}) for five-point tensor integrals.
The scalar integrals are calculated as in Refs.~\cite{'tHooft:1978xw, Dittmaier:2003bc, Nhung:2009pm, Denner:2010tr}.
The UV divergences of the loop integrals are regularized
using dimensional regularization (DR)~\cite{'tHooft:1972fi}.
Since the light quarks are approximated as massless,
their mass counterterms vanish.

The real-emission processes are classified into the gluon-radiated processes
\begin{align}
\bar{q} + q & \to W^+ + W^- + Z + g
\label{qqWWZ_qcd_rad}
\end{align}
and the gluon-induced processes
\begin{align}
q + g & \to W^+ + W^- + Z + q,\crn
\bar{q} + g & \to W^+ + W^- + Z + \bar{q}.
\label{qqWWZ_qcd_ind}
\end{align}
Both the virtual and real corrections are
separately IR divergent. These divergences cancel in the sum for
infrared-safe observables such as the total cross section and
kinematic distributions of massive gauge bosons.
The IR singularities are treated using the DR
and mass regularization (MR) schemes (see also \sect{xsection_had}).
MR method uses a common mass regulator for the light
fermions (all but the top quark) and a fictitious gluon mass.
The results of two schemes are in agreement.

Moreover, we apply the Catani-Seymour dipole subtraction
algorithm~\cite{Catani:1996vz} to combine the virtual and the
real contributions. We use the same notations as in
\bib{Catani:1996vz} with the DR method and define the various NLO QCD corrections as
follows,
\bea
\sigma_\text{QCD-virt} &=& \int \dd x_1\dd x_2[\bar{q}_\text{NLO}(x_1,
\mu_F)q_\text{NLO}(x_2, \mu_F)\hsigma^{\bar{q}q\to
  W^+W^-Z}_\text{QCD-virt}+(1\leftrightarrow 2)],\crn
\hsigma^{\bar{q}q\to W^+W^-Z}_\text{QCD-virt} &=&
\hsigma^{\bar{q}q\to W^+W^-Z}_\text{QCD-loop} +
\hsigma^{\bar{q}q\to W^+W^-Z}_\text{QCD-I},\,
\label{xsection_virt_qcd}
\eea
where $\hsigma^{\bar{q}q\to W^+W^-Z}_\text{QCD-loop}$
includes only loop diagrams and $\hsigma^{\bar{q}q\to
  W^+W^-Z}_\text{QCD-I}$ is the I-operator contribution as defined
in \bib{Catani:1996vz}. It is noted that
$\hsigma^{\bar{q}q\to W^+W^-Z}_\text{QCD-virt}$ is UV and
IR finite. The gluon-radiated and gluon-induced contributions read
\bea
\sigma_\text{g-rad} &=& \int \dd x_1\dd x_2[\bar{q}_\text{NLO}(x_1,
\mu_F)q_\text{NLO}(x_2, \mu_F)
\left(\hsigma^{\bar{q}q\rightarrow W^+W^-Zg} -
  \hsigma^{\bar{q}q\rightarrow
    W^+W^-Z}_\text{QCD-I}\right)+(1\leftrightarrow 2)],\crn
\sigma_\text{g-ind} &=& \int \dd x_1\dd x_2[q_\text{NLO}(x_1,
\mu_F)g_\text{NLO}(x_2, \mu_F)
\hsigma^{qg\rightarrow W^+W^-Zq} + (1\leftrightarrow
2)].
\label{xsection_real_qcd}
\eea
These contributions are also IR finite because the collinear
divergences occurring at partonic level are absorbed into the quark
PDFs.

%%%%%%%%%%%%%%%%%%%%%%%%%%%%%%%%%
\subsection{NLO EW corrections} %
\begin{figure}[h]
  \centering
   \includegraphics[width=01\textwidth,height=0.60\textwidth]
{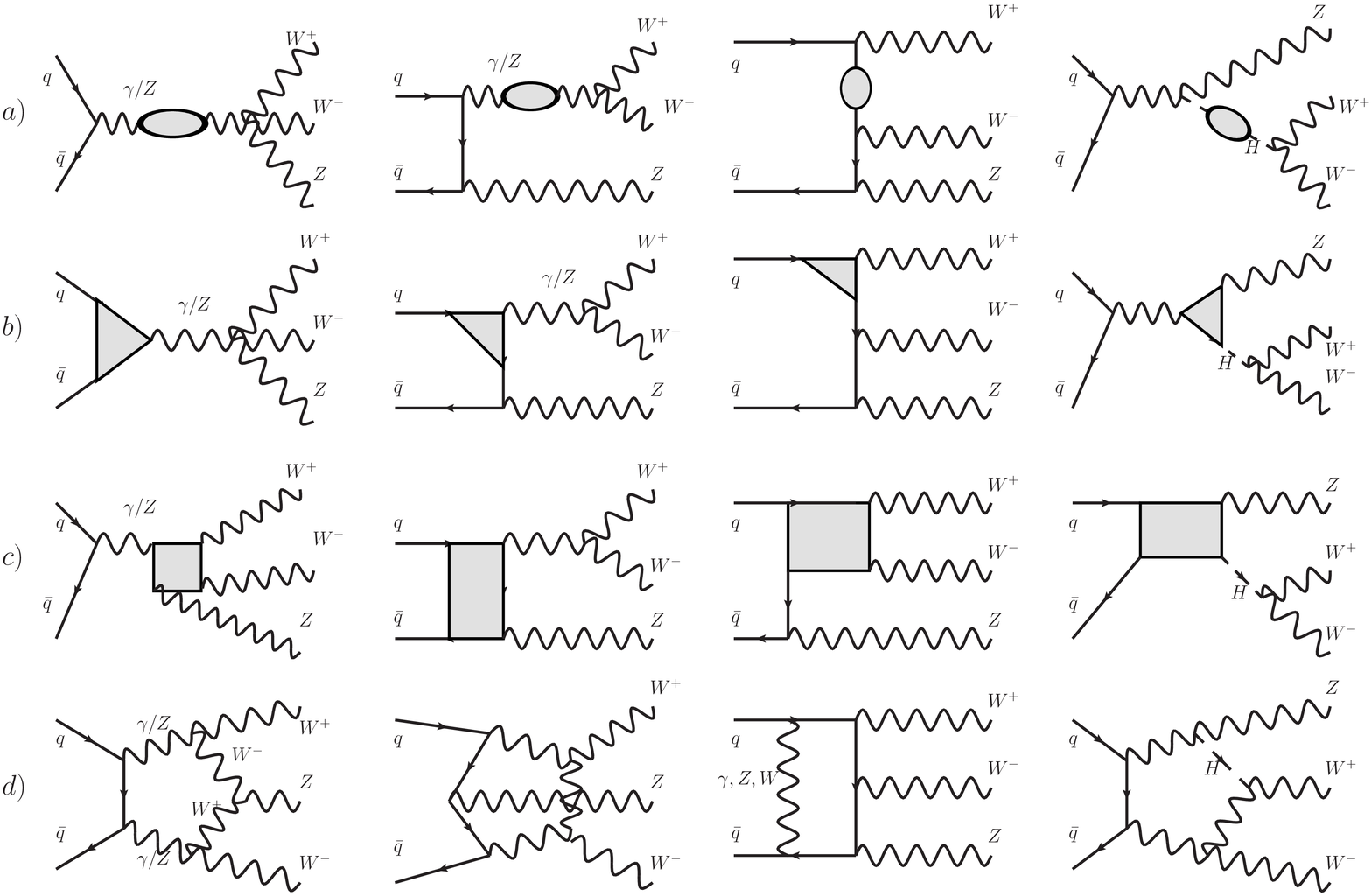}
  \caption{Representative sets of self-energy, vertex, box and pentagon diagrams. The shaded regions are
the one-particle irreducible two-, three- and four-point vertices including possible counterterms.}
\label{qqWWZ_EW_diag}
\end{figure}

\label{cal_ew}                  %
%%%%%%%%%%%%%%%%%%%%%%%%%%%%%%%%%
The NLO EW contribution also includes the virtual and real corrections.
Compared to the QCD case, the virtual EW contribution is much more complicated.
The one-loop Feynman diagrams contain extra bosons ($\gamma$, $Z$, $W^\pm$ or $H$) in the
loops or a fermion loop. The presence of fermion loops with $\gamma_5$ requires that all leptons
and quarks contribution must be included to cancel the anomaly. For illustration, representative sets
of two-, three-, four- and five-point vertices are shown in \fig{qqWWZ_EW_diag}($a,b,c,d$), respectively.
As in the QCD case, the NLO EW corrections involve also five-point tensor integrals up to rank four, see the third Feynman
graph in \fig{qqWWZ_EW_diag}(d). The one-loop integrals are calculated using the same method as in the QCD case.

We now discuss the issue of renormalization to deal with UV divergences.
Renormalization of the electric coupling,
the gauge boson masses, the Higgs mass and the external wave functions
are performed.
We adopt the on-shell renormalization scheme (see \cite{Aoki:1982ed, Denner:1991kt, Belanger:2003sd}) with a little modification
of the electric charge renormalization constant as specified in the following.
The bare electric charge is related to the renormalized one through a renormalization constant as
$e_0 = e (1+ \de Z_e)$. The on-shell condition for the photon-fermion-fermion vertex at the Thomson limit ($k \to 0$)
together with the Ward identity lead to a relation of the electric charge renormalization constant with the
photon wave-function renormalization constants as
\bea \de Z_e^{\alpha(0)} = -\fr 12 \de Z_{AA} - \fr{s_W}{2c_W} \de Z_{ZA},\quad  \de Z_{AA} = -\fr{\pa \Si_{T}^{AA}(k^2)}{\pa k^2}\bigg{|}_{k^2\to 0},\quad
 \de Z_{ZA}=2\fr{\Si_{T}^{AZ}(0)}{M_Z^2},\eea
where $s_W=\sin\theta_W$, $c_W=\cos\theta_W$ with $\theta_W$ being the weak mixing angle
and $\Si_{T}^{XY}(k^2)$ is the transverse part of the unrenormalized self-energy of
the $X\to Y$ transition at momentum squared $k^2$.
The derivative of the photon self-energy in the vanishing momentum limit
introduces a logarithm, $\log(m_f^2/q^2)$,
with the fermion mass $m_f$ and a typical energy scale $q$ of the hard process.
This logarithm becomes problematic for the light quarks
since their masses are not well measured.
For the process with tree-level amplitude proportional to ${\cal O}(e^n)$
and including $n$ external photons,
the NLO EW correction is free of those logarithms
due to the cancellation between those from the vertex
counterterms and the one arising from
external photon wave-function counterterms.
This can also be seen by the observation that
all vertices at tree level involve a real photon, hence the
running of the electric charge is absent.
The logarithmic correction remains if the
number of external photons is less than $n$ as in our process.
This correction being universal can be absorbed into the running of $\alpha$ using
$\alpha(M_Z^2)$ or using the $G_\mu$-scheme with
\be \al_{G_\mu} = \fr{\sqrt{2}M_W^2 G_{\mu}}{\pi}\braket{1-\fr{M_W^2}{M_Z^2}}, \label{eq_alfa_Gmu}\ee
where $G_\mu$ is the Fermi constant. We will choose the latter and use $\al_{G_\mu}$ as an input parameter.
By considering one-loop EW corrections to the muon decay, one finds the quantity $\De r$ \cite{Sirlin:1980nh}
\bea
\De r &=&  -\de Z_{AA} -\fr{c_W^2}{s_W^2}\braket{\fr{\Si_T^{ZZ}(M_Z^2)}{M_Z^2} -\fr{\Si_T^{WW}(M_W^2)}{M_W^2}}  +\fr{\Si_T^{WW}(0) -\Si_T^{WW}(M_W^2)}{M_W^2}\crn
      && + 2\fr{c_W}{s_w}\fr{\Si_T^{AZ}(0)}{M_Z^2}+\fr{\al}{4\pi s_W^2}\braket{6 +\fr{7-4s_W^2}{2s_W^2}\log c_W^2}.
\eea
This leads to the relation
\bea
\alpha_{G_\mu} = \fr{\alpha(0)}{1 - \De r}.
\label{alfa_running}
\eea
To avoid double counting at NLO EW corrections, the electric charge renormalization constant is modified as
\be  \de Z_e^{G_\mu} = -\fr 12 \de Z_{AA} - \fr{s_W}{2c_W} \de Z_{ZA} -\fr 12 \De r,\ee
which leads to the cancellation of $\de Z_{AA}$ in $\de Z_e^{G_\mu}$.
As long as no external photon appears at tree-level
the NLO EW corrections will be insensitive to the light fermion masses
with the above modification of the electric charge renormalization constant.
Therefore,
the light quark masses are set to zero everywhere unless their masses are
used as regulators of collinear singularities.
To summarize the use of $\al_{G_\mu}$ scheme in our calculation,
the LO partonic cross sections are of ${\cal O}(\al^3_{G_\mu})$,
the NLO QCD ones are of  ${\cal O}(\al^3_{G_\mu}\al_s)$.
For the NLO EW corrections, they contain real-photon emission contributions
where the coupling of a real photon should be $\alpha(0)$.
The NLO EW cross section is therefore of ${\cal O }(\al^3_{G_\mu}\al(0))$.
For the $\gamma\gamma \to WWZ$ process, the tree-level cross section is of 
${\cal O }(\al_{G_\mu}\al(0)^2)$. 

\begin{figure}[h]
  \centering
   \includegraphics[width=01\textwidth,height=0.360\textwidth]
{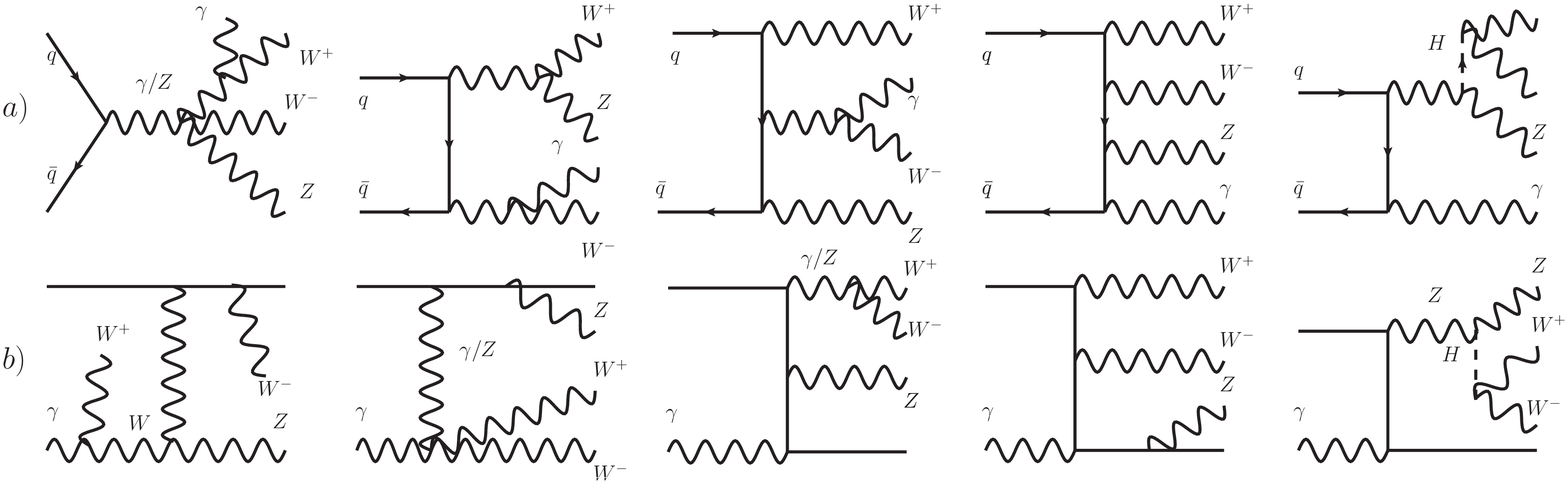}
  \caption{Representative Feynman diagrams for the real photon radiation a) and the photon induced subprocesses b).
The solid straight lines stand for the (anti-)quarks.}
\label{photonic_diag}
\end{figure}

The real-emission corrections contain an extra photon in the external state.
Similar to the QCD case, we have the photon-radiated processes
\bea
\bar{q} + q & \to W^+ + W^- + Z + \gamma
\label{qqWWZ_ew_real}
\eea
and the photon-induced processes
\bea
\begin{aligned}
q + \gamma & \to W^+ + W^- + Z + q,\\
\bar{q} + \gamma & \to W^+ + W^- + Z + \bar{q},
\end{aligned}
\eea
whose Feynman diagrams are shown in \fig{photonic_diag}(a,b), respectively.
Compared to the real-gluon emission correction,
the IR-singularity structure in the photonic correction is much more complicated.
In the real-photon radiation case, the singularities
arise from two types of splittings: $q\to q^* \gamma$ and $W^*\to W\gamma$.
The former gives rise to both soft and collinear divergences while the latter introduces only soft divergences via
interference effects.
For the photon-induced subprocesses, there occur only collinear divergences
arising from the following splittings: $q\to q \gamma^*$ and $\gamma \to q^* \bar{q}$.

In order to deal with those IR divergences and to combine the real-emission
and virtual corrections, we will follow the convention of
\bib{Dittmaier:1999mb}. We use the MR method to
regularize IR divergences.
For the $\bar{q}q\to W^+W^-Z$ processes, the correction
$\sigma_\text{EW-virt}$ is, similarly to the QCD case, given as in
\eq{xsection_virt_qcd}, but the I-operator contribution
$\hsigma^{\bar{q}q \to W^+W^-Z}_\text{EW-I}$ is now
defined as the endpoint contribution of \bib{Dittmaier:1999mb}. The
photon-radiated and photon-induced contributions read
\begin{align}
\sigma_{\gamma\text{-rad}} &= \int \dd x_1\dd
x_2[\bar{q}_\text{NLO}(x_1, \mu_F)q_\text{NLO}(x_2, \mu_F)
\left(\hsigma^{\bar{q}q\rightarrow W^+W^-Z\gamma} -
  \hsigma^{\bar{q}q\rightarrow
    W^+W^-Z}_\text{EW-I}\right)+(1\leftrightarrow 2)],\crn
\sigma_{\gamma\text{-ind}} &= \int \dd x_1\dd x_2[q_\text{NLO}(x_1,
\mu_F)\gamma_\text{NLO}(x_2, \mu_F)
\hsigma^{q\gamma\rightarrow W^+W^-Zq} + (1\leftrightarrow
2)].
\label{xsection_real_ew}
\end{align}
For EW corrections, we use $f_\text{NLO}(x, \mu_F) = f_\text{LO}(x,
\mu_F)$ for $f=q,\bar{q},\gamma$ as will be discussed in \sect{xsection_had}.
Moreover, the collinear
divergences occurring at the partonic level in the photon-radiated and
photon-induced contributions are absorbed into the (anti-)quark and
photon PDFs using the DIS factorization scheme as described in
\sect{xsection_had}.

The aforementioned method has been implemented in different computer
codes, using the {\texttt{FORTRAN77}} and {\texttt{C++}} programming
languages. The helicity amplitudes are generated using
{\texttt{FeynArts-3.4}}~\cite{Hahn:2000kx} and
{\texttt{FormCalc-6.0}}~\cite{Hahn:1998yk} as well as
{\texttt{HELAS}}~\cite{Murayama:1992gi,Alwall:2007st}. The scalar and tensor
one-loop integrals in one code are evaluated with the in-house library
{\texttt{LoopInts}}. This library has an option to use quadruple precision, on the
fly, when numerical instabilities are detected. We have observed that
the numerical integration of the virtual corrections, in particular for
the EW case, shows numerical instabilities. One of our solutions to this problem
is described as follows. When using the MR method, the small mass regulators are neglected
as much as possible for IR-safe one-loop integrals. This has to be consistently done
from the top level of tensor coefficients to the bottom level of scalar integrals to ensure
a regular behavior of the tensor coefficients in the limit of vanishing Gram
determinant ($\text{det}(2p_ip_j)$ with $p_i$ being external momenta). After this step, the
Gram determinant is checked for $N$-point tensor coefficients ($N=3,4$),
and if it is small enough, \ie\
\bea
\fr{\text{det}(2p_ip_j)}{(2p^2_\text{max})^{N-1}} < 10^{-3}, \quad i,j = 1,\ldots, N-1,
\eea
where $p^2_\text{max}$ is the maximum external mass of a triangle or box diagram, then all those
tensor coefficients are calculated with quadruple precision. Otherwise double precision is used.
For five-point tensor coefficients, we use the method of \bib{Denner:2005nn} to
avoid the small Gram determinant problem.
Moreover, the real corrections have been checked
by comparing the results of the dipole-subtraction method with those
of the phase-space slicing method~\cite{Baur:1998kt}.

%%%%%%%%%%%%%%%%%%%%%%%%%%%%%%%%%
\subsection{Hadronic cross section}%
\label{xsection_had}            %
%%%%%%%%%%%%%%%%%%%%%%%%%%%%%%%%%
The LO hadronic cross section is given by
\bea
\sigma_\text{LO}= \int \dd x_1\dd x_2[\bar{q}_\text{LO}(x_1, \mu_F)q_\text{LO}(x_2, \mu_F)\hsigma^{\bar{q}q}_{LO}(\alpha^3)+(1\leftrightarrow 2)],
\label{xsection_LO}
\eea
where $q$ and $\bar{q}$ are LO parton distribution
functions  of the light quarks in the proton at momentum fraction $x$ and factorization scale $\mu_F$.
The bottom-quark contribution $\sigma_{\bar{b}b}$ is calculated in the same way. The top-quark contribution is neglected and
the photon contribution reads
\bea
\sigma_{\gamma\gamma}= \int \dd x_1\dd x_2[\gamma(x_1, \mu_F)\gamma(x_2, \mu_F)\hsigma_{\gamma\gamma}(\alpha^3)],
\label{xsection_gam}
\eea
where the photon PDF is given by the code MRSTQED2004~\cite{Martin:2004dh} as discussed below.

The NLO hadronic cross section is defined as follows:
\bea
\sigma_\text{NLO} = \sigma^{\bar{q}q}_\text{QCD}(\alpha^3,\alpha^3\alpha_s)
+ \Delta\sigma^{\bar{q}q}_\text{EW}(\alpha^4) + \sigma_{\bar{b}b}(\alpha^3) + \sigma_{\gamma\gamma}(\alpha^3) ,
\label{xsection_NLO}
\eea
where the first term including the
tree-level and NLO QCD corrections is calculated with NLO PDFs, the second term is
the NLO EW correction.

We now discuss the issue of PDFs. Ideally, we would choose a NLO PDF set including QCD and EW corrections for the NLO results.
However, there exists at the present no PDF set with NLO EW corrections.
The leading EW contribution is included in the MRSTQED2004 set, and very
recently also in the NNPDF set~\cite{Carrazza:2013bra}.
In our case, since the $\bar{q}q$ contribution is dominant we will use the more reliable MSTW2008 PDF set \cite{Martin:2009iq}
everywhere for initial quarks. This set includes only QCD corrections.
The photon PDF is needed for the LO $\gamma\gamma$ and the EW real corrections
with photon in the initial state. For these contributions, we get the photon PDF from the MRSTQED2004 set.
For NLO QCD corrections, since the PDFs are defined in the $\MSb$ factorization scheme the one-loop calculation in
\sect{cal_qcd} is also done in this scheme.
For NLO EW corrections, \ie\ the second term in \eq{xsection_NLO}, we use the LO PDF set and the calculation
is done by assuming the $\DIS$ factorization scheme. We can also take the $\MSb$ scheme as in the QCD case, but
there is really no justification for either choice since the quark PDFs include no EW corrections. We choose the
$\DIS$ scheme because it is usually used for NLO EW corrections (see \eg\ \cite{Diener:2005me}).
Accordingly, the PDF counterterms which appear in the real corrections are defined as follows, here $q$ stands for both
quarks and anti-quarks, in mass regularization (MR)
\bea
\delta^\MR q(x, \mu_{\text F}^2)& =&  -\fr{\al_s C_F}{2\pi}
\int_x^1\fr{dz}{z} q\left(\fr xz, \mu_F^2\right)
 \bigg\{ \ln\bpmatrix\fr{\mu_F^2}{m_q^2}\epmatrix
[P_{qq}(z)]_+ + P_{qq}^\text{reg}(z) + C_{qq}^{\MSb}(z) \bigg\} \crn
&& -\fr{\al Q_q^2}{2\pi}
\int_x^1\fr{dz}{z} q\left(\fr xz, \mu_F^2\right)
 \bigg\{ \ln\bpmatrix\fr{\mu_F^2}{m_q^2}\epmatrix
[P_{qq}(z)]_+ + P_{qq}^\text{reg}(z) + C_{qq}^{\DIS}(z) \bigg\} \crn
&&-\, \fr{\al_sT_F }{2\pi}\int_x^1\fr{dz}{z} g\left(\fr xz, \mu_F^2\right)
\bigg[ \ln\bpmatrix \fr{\mu_F^2}{m_q^2}\epmatrix P_{gq} + C_{gq}^{\MSb}(z) \bigg]\crn
&&-\, \fr{3\al Q_q^2}{2\pi}\int_x^1\fr{dz}{z} \gamma\left(\fr xz, \mu_F^2\right)
\bigg[ \ln\bpmatrix \fr{\mu_F^2}{m_q^2}\epmatrix P_{\gamma q} + C_{\gamma q}^{\DIS}(z) \bigg],
\label{q_redifined_mass}\\
%%%
\delta^\MR \gamma(x, \mu_{\text F}^2) &=& -\, \fr{\al}{2\pi}\sum_{q} Q_q^2 \int_x^1\fr{dz}{z} q\left(\fr xz, \mu_F^2\right)
\bigg[ \ln\bpmatrix \fr{\mu_F^2}{m_q^2}\epmatrix P_{q\gamma} + P_{q\gamma}^\text{reg}(z) + C_{q\gamma}^{\DIS}(z) \bigg]
,
\label{y_redifined_mass}
\eea
with $C_F = 4/3$, $T_F=1/2$ and (see \eg\ \cite{Dittmaier:2009cr})
\bea
P_{qq}^\text{reg}(z) &=& -[P_{qq}(z)(2\ln(1-z) +1)]_+ , \crn
P_{q\gamma}^\text{reg}(z) &=& -P_{q\gamma}(z)(2\ln z + 1) .
\label{P_reg}
\eea
The corresponding DR counterterms, $\delta^\DR q(x, \mu_{\text F}^2)$ and $\delta^\DR \gamma(x, \mu_{\text F}^2)$,
are obtained from Eqs.~(\ref{q_redifined_mass},\ref{y_redifined_mass}) using the following rules
\bea
\log(m_q^2) \to \fr{1}{\eps} - \gamma_\text{E} + \log(4\pi\mu^2), \quad  P_{qq}^\text{reg}(z) \to 0, \quad P_{q\gamma}^\text{reg}(z) \to 0,
\label{rule_redifined_dim}
\eea
where we have used $D=4-2\eps$, $\gamma_\text{E}$ is Euler's constant and $\mu$ is the usual mass-dimension parameter in DR.
This replacement rule agrees with the standard definition in \cite{Catani:1996vz} for DR. Moreover, we have explicitly checked
\eq{rule_redifined_dim} by verifying numerically for various processes~\cite{Baglio:2013toa} that the results obtained using MR agree with the DR ones.
The gluon PDF does not occur at LO in our calculation, therefore its counterterm does not appear at NLO.

The splitting functions are given by
\bea P_{qq}(z) = \fr{1+z^2}{1-z}, \quad
P_{gq}(z) = P_{\gamma q}(z) = z^2 + (1-z)^2, \quad P_{q\gamma}(z) = \fr{1+(1-z)^2}{z},
\eea
and the $[\ldots]_{+}$ prescription is understood in the usual way,
\bea
\int_0^1dz [g(z)]_{+} f(z)=\int_0^1dz\, g(z) [f(z)-f(1)].
\eea
The factorization schemes are specified by \cite{Catani:1996vz}
\bea
C_{qq}^{\MSb}(z) &=& C_{gq}^{\MSb}(z) = 0,\crn
C_{qq}^{\DIS}(z) &=& \left[P_{qq}(z)\left(\ln(\fr{1-z}{z}) - \fr{3}{4}\right) + \fr{9+5z}{4} \right]_+ ,\crn
C_{\gamma q}^{\DIS}(z) &=& P_{\gamma q}\ln(\fr{1-z}{z}) - 8z^2 + 8z - 1,\; C_{q\gamma}^{\DIS}(z) = -C_{qq}^{\DIS}(z).
\eea

\begin{comment}
We have, at NLO QCD,
\bea
\sigma^{\bar{q}q}_\text{QCD}= \sum_{\bar{q},q}
\int \dd x_1\dd x_2[F_{\bar{q}}^\text{NLO}(x_1, \mu_F)F_{q}^\text{NLO}(x_2, \mu_F)
\hsigma^{\bar{q}q}_{QCD}(\alpha^3,\alpha^3\alpha_s,\mu_R)+(1\leftrightarrow 2)],
\eea
and at NLO EW
\bea
\sigma^{\bar{q}q}_\text{EW}= \sum_{\bar{q},q}
\int \dd x_1\dd x_2[F_{\bar{q}}^\text{LO}(x_1, \mu_F)F_{q}^\text{LO}(x_2, \mu_F)
\hsigma^{\bar{q}q}_{EW}(\alpha^3,\alpha^4,\mu_R)+(1\leftrightarrow 2)].
\eea
\end{comment}

%%%%%%%%%%%%%%%%%%%%%%%%%%%%%%%
\section{Numerical results}   %
\label{sect_result}           %
%%%%%%%%%%%%%%%%%%%%%%%%%%%%%%%

We use the following set of input parameters \cite{Beringer:1900zz,:2012gk,:2012gu},
\begin{equation}
\begin{aligned}
G_\mu &= 1.16637\times 10^{-5}\gev^{-2}, \hs \alpha(0)=1/137.035999679 , \hs \alpha_s(M_Z) = 0.12018, \\
M_{W} &= 80.385\gev, \hs M_Z= 91.1876\gev, \hs m_t=173.5\gev, \hs M_H = 125\gev,
\end{aligned}
\end{equation}
where the strong coupling constant $\alpha_s(M_Z)$ occurs only in the NLO QCD corrections and
is determined from the NLO MSTW2008 PDF set with five quark 
flavors as discussed in \sect{xsection_had}. The
Cabibbo-Kobayashi-Maskawa matrix is set to be diagonal.
The masses of the light quarks, \ie\ all but the top mass,
are approximated as zero.
This is justified because our results are
insensitive to those small masses. As argued in \sect{cal_ew}, the
NLO EW corrections are proportional
to $\alpha_{G_{\mu}}^3\alpha(0)$ where $\alpha_{G_{\mu}}$ is calculated
as in \eq{eq_alfa_Gmu}.
We also use $\alpha(0)$ as an input parameter because the relation~(\ref{alfa_running})
involving the hadronic contribution to the photon self-energy at low energy is not reliable
and hence we do not use it to calculate $\alpha(0)$. The $\gamma\gamma$ contribution is 
of ${\cal O }(\al_{G_\mu}\al(0)^2)$. In the following we present the results for the LHC at $14\tev$.

\subsection{Total cross section}
\label{total_Xsection}
\begin{figure}[t]
 \begin{center}
\includegraphics[width=0.6\textwidth]{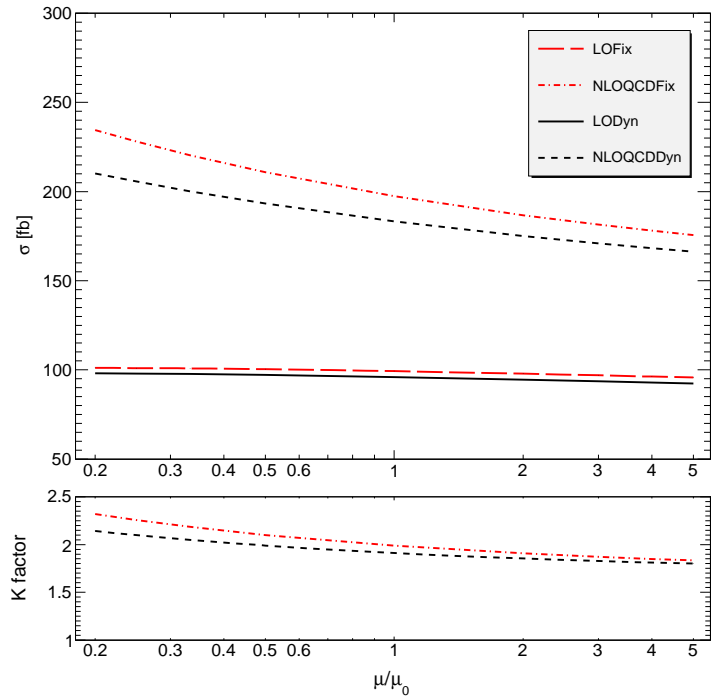}
\caption{Total cross sections and $K$ factor (defined in the text) as functions of the
scale $\mu = \mu_F = \mu_R$.}
\label{scale}
\end{center}
\end{figure}

The NLO results depend on the renormalization scale $\mu_R$ and factorization scale $\mu_F$, which are
arbitrary parameters. $\mu_R$ occurs via the strong coupling constant and explicitly in the
virtual QCD amplitude. The virtual EW amplitude does not introduce any $\mu_R$ dependence since it
is calculated using the OS renormalization scheme. $\mu_F$ occurs via the PDFs and
explicitly in the real QCD amplitude. The EW factorization scale dependence is much smaller than
the QCD one, hence can be neglected. The scales $\mu_F$ and $\mu_R$ are hereafter meant to be of QCD origin.
For simplicity they will be set equal and be referred to as the scale $\mu$.

In \fig{scale} we show the LO and the NLO QCD total cross sections as functions of $\mu$ varied
around the center scale $\mu_0$ for two cases: a fixed scale with
$\mu_0=2M_W+M_Z$ and a dynamic
scale $\mu_0=M_{WWZ}$, the invariant mass of triple-boson system.
The $K$ factor, defined as the ratio of the NLO to the LO results, is in the lower panel.
We observe that the NLO QCD correction is about $100\%$ and the scale uncertainty does
not give a good estimate of the higher-order contribution. The fixed-scale results
are similar to the dynamic ones for both the total cross section and the distributions we
have studied. The small $\mu$ dependence of the LO total cross section can be explained
as follows. The $M_{WWZ}$ distribution is maximal near the threshold, at $M_{WWZ}^\text{max}\approx 400\gev$.
This corresponds to $\sqrt{x_1x_2} = M_{WWZ}^\text{max}/\sqrt{s}=0.03$ for $\sqrt{s}=14\tev$.
The rapidity $WWZ$ distribution is maximal at $y_{WWZ}^\text{max}=0$, which means $x_1 = x_2$.
Thus, the main contribution to the total cross section comes from the
region $x_1 = x_2 = 0.03$ where the PDFs have a small factorization scale
dependence. The same argument holds for the NLO results, hence
the scale dependence at NLO is given mainly by the renormalization scale.
Some values of the total cross section corresponding to \fig{scale} are
given in \tab{table_scale_14}.
\begin{table}[h]
 \begin{footnotesize}
 \bc
 \caption{\label{table_scale_14}{Total cross section (in fb) shown in
     \fig{scale} as function of the scale $\mu = \mu_F = \mu_R$.}}
\vspace*{0.5cm}
\begin{tabular}{|l|c | r@{.}l | r@{.}l | r@{.}l | r@{.}l |}
 \hline
\multicolumn{2}{|c|}{}&\multicolumn{4}{c|}{Fixed scale} & \multicolumn{4}{c|}{Dynamic scale}\\
\hline
\multicolumn{2}{|c|}{$\mu$} &\multicolumn{2}{c|}{ LO}
&\multicolumn{2}{c|}{ NLO QCD}
&\multicolumn{2}{c|}{ LO}
&\multicolumn{2}{c|}{ NLO QCD}
\\
\hline
\multicolumn{2}{|l|}{$\mu_0/4$}&   101&02(2)  & 227&94(4) & 97&98(2) & 205&57(4) \\
\multicolumn{2}{|l|}{$\mu_0/2$}&  100&39(2)  & 210&76(4) & 97&11(2) & 193&25(3) \\
\multicolumn{2}{|l|}{$\mu_0$}&  99&29(2)  & 197&41(4) & 95&91(2) & 183&31(3) \\
\multicolumn{2}{|l|}{$2\mu_0$}&  97&87(2)  & 186&70(3) & 94&48(2) & 175&11(3) \\
\multicolumn{2}{|l|}{$4\mu_0$}&  96&25(2)  & 178&01(3) & 92&91(2) & 168&26(3) \\
\hline
\end{tabular}\ec
 \end{footnotesize}
\end{table}
The NLO QCD corrections have been calculated by two groups \cite{Hankele:2007sb, Hankele:2009phd} and \cite{Binoth:2008kt}.
It is stated in \cite{Hankele:2009phd} that the two results agree for the case of no Higgs contribution.
We have made a comparison with those groups, using the same input parameters as in \cite{Binoth:2008kt},
and found very good agreement at LO.
The agreement at NLO is at the level of $1.5\%$.

%%%%%%%%%
\begin{table}[h]
 \begin{footnotesize}
 \bc
 \caption{\label{table_total_14}{
 Total cross section in fb for $pp \to W^+ W^-Z$ including the QCD NLO
and EW NLO corrections   at $\sqrt{s}= 14\tev$ for fixed scale $\mu_F = \mu_R = 2M_W + M_Z$ and
dynamic scale $\mu_F = \mu_R = M_{WWZ}$.
The numbers in brackets show the integration uncertainty in the last digit if  they are significant.}}
\vspace*{0.5cm}
\begin{tabular}{|l|c | r@{.}l | r@{.}l | r@{.}l | r@{.}l |}
 \hline
\multicolumn{2}{|c|}{}&\multicolumn{4}{c|}{Fixed scale} & \multicolumn{4}{c|}{Dynamic scale}\\
\hline
\multicolumn{2}{|c|}{} &\multicolumn{2}{c|}{ $\si [fb]$}
&\multicolumn{2}{c|}{ $\de [\%] $}
&\multicolumn{2}{c|}{$\si [fb]$}
&\multicolumn{2}{c|}{$\de [\%] $}
\\
\hline
\multicolumn{2}{|l|}{LO}            &  99&29(2)  &  .&.  &  95&91(2)  & .&.    \\
\multicolumn{2}{|l|}{$\bar{b}b$}     &  2&4173    & 2&4   &   2&6915   & 2&8         \\
\multicolumn{2}{|l|}{$\ga\ga$ }     &  4&852     & 4&9   &   5&559    & 5&8  \\
\hline
\multirow{2}{*}{$\De_\text{QCD}$}& $q\bar q$        &  48&83(3)  & 49&2 &  53&33(3) & 55&6 \\
                               & $q g, \bar q g$  &  49&29(1)  &  49&6 &  34&07(1) & 35&5 \\
\hline
\multirow{2}{*}{$\De_\text{EW}$}& $q\bar q$           &   -8&74(1) & -8&8 &  -8&05(1) & -8&4 \\
                           & $q \ga, \bar q \ga$ &   6&81(1)  & 6&8  &  5&854(9) &  6&1 \\
\hline
\multicolumn{2}{|l|}{$\De_\text{NLO}$}       &  103&46(4) & 104&2 &  93&46(4) & 97&4 \\
\hline
\end{tabular}\ec
 \end{footnotesize}
\end{table}
We now include the NLO EW corrections as well the LO $\bar{b}b$ and $\gamma\gamma$
contributions. They are shown in \tab{table_total_14} for the fixed and
dynamic scale choices. In this table and the following discussions
the relative corrections are normalized to $\sigma_\text{LO}$ defined in \eq{xsection_LO}.
The correction coming from $\bar{b}b$ initial state is less than $3\%$ while
the $\gamma\gamma$ one is about two times lager. 
For the dynamic scale choice, if $\mu_F = M_{WWZ}$ is outside the allowed
energy range of the MRSTQED2004 code, namely $\mu^2_F>  10^7\gev^2$, then
the photon PDF is set to zero. The impact of this cut should be very small
since the contribution from that phase-space region is suppressed. 
The study of the EW correction to $\ga\ga\to W^+W^-$ in \bib{Baglio:2013toa} gives, for the total
cross section, a per mille correction on top of the LO $\ga\ga$ contribution.
We also expect the same effect for $\ga\ga \to W^+W^-Z$, hence NLO EW corrections to this
subprocess are neglected.

In \tab{table_total_14} and also in \sect{sect_dist}
we also show several subcorrections as defined in \sect{cal_qcd}
for the QCD case and in \sect{cal_ew}
for the EW case.
For the QCD correction, we have:
the PDF correction coming from the difference between the NLO and LO PDFs,
the gluon-radiated correction, the gluon-induced correction
and the virtual correction, as defined in \sect{cal_qcd}.
The PDF, virtual and gluon-radiated corrections are combined in the entry $q\bar{q}$
in \tab{table_total_14}, but they are separately shown in \sect{sect_dist}.
Similarly, the EW correction is also separated into
the photon-radiated, photon-induced and virtual corrections.
The PDF correction vanishes because the LO PDFs are used for the EW corrections.
The virtual and photon-radiated corrections are combined in the entry $q\bar{q}$
in \tab{table_total_14}, but they are separately shown in \sect{sect_dist}.
In the case of the QCD corrections the
$q\bar q$ and gluon-induced contributions are of the same order of magnitude
and have the same positive sign.
In contrast, the two contributions in the EW correction have opposite signs.
This makes the total NLO EW correction about $-2\%$.

We close this subsection with some comments on the single-top contribution.
If one considers the NLO QCD corrections to $\bar{b}b \to W^+W^-Z$ channel, there is
a large contribution from the gluon-induced process $b g \to W^+W^-Z b$ due
to the mechanism $b g \to W^-Z t (t \to W^+ b)$ with an intermediate on-shell
top quark. This large $WZt$ production mode,
being a part of the singe-top background, should be excluded and our
main concern is the interference between this mechanism and the genuine $WWZ$ channel
without the on-shell top quark. As in the $W^+W^-$ case~\cite{Baglio:2013toa}, this
interference effect is expected to be negligible. We therefore neglect the
NLO QCD corrections to $\bar{b}b \to W^+W^-Z$ subprocess.

\subsection{Distributions}
\label{sect_dist}
We do not observe any important difference
between the fixed scale and dynamic scale results for various distributions.
We therefore show only some representative distributions with the fixed scale choice.

\begin{figure}[h]
 \begin{center}
\includegraphics[width=0.32\textwidth]{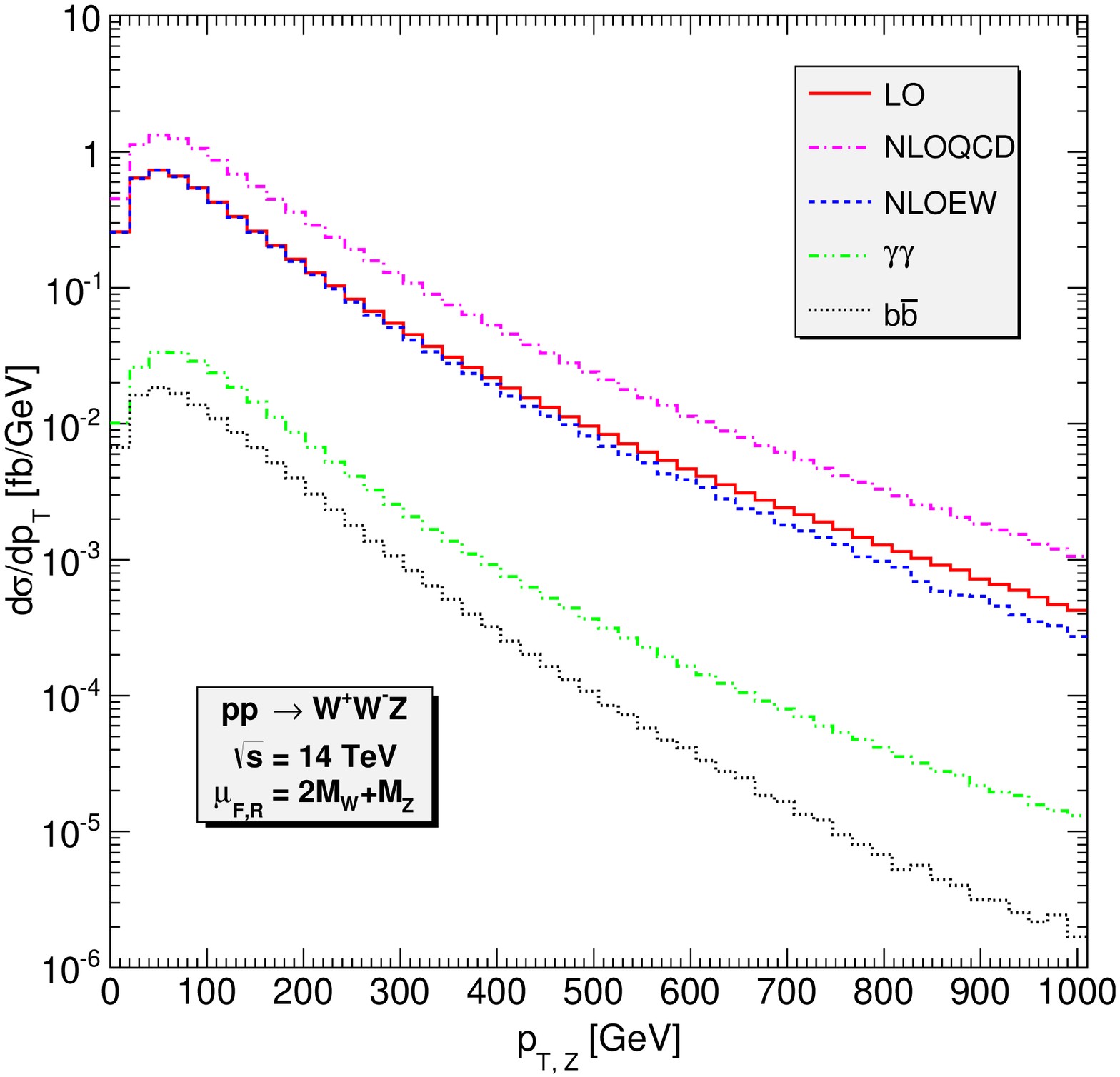}
\includegraphics[width=0.32\textwidth]{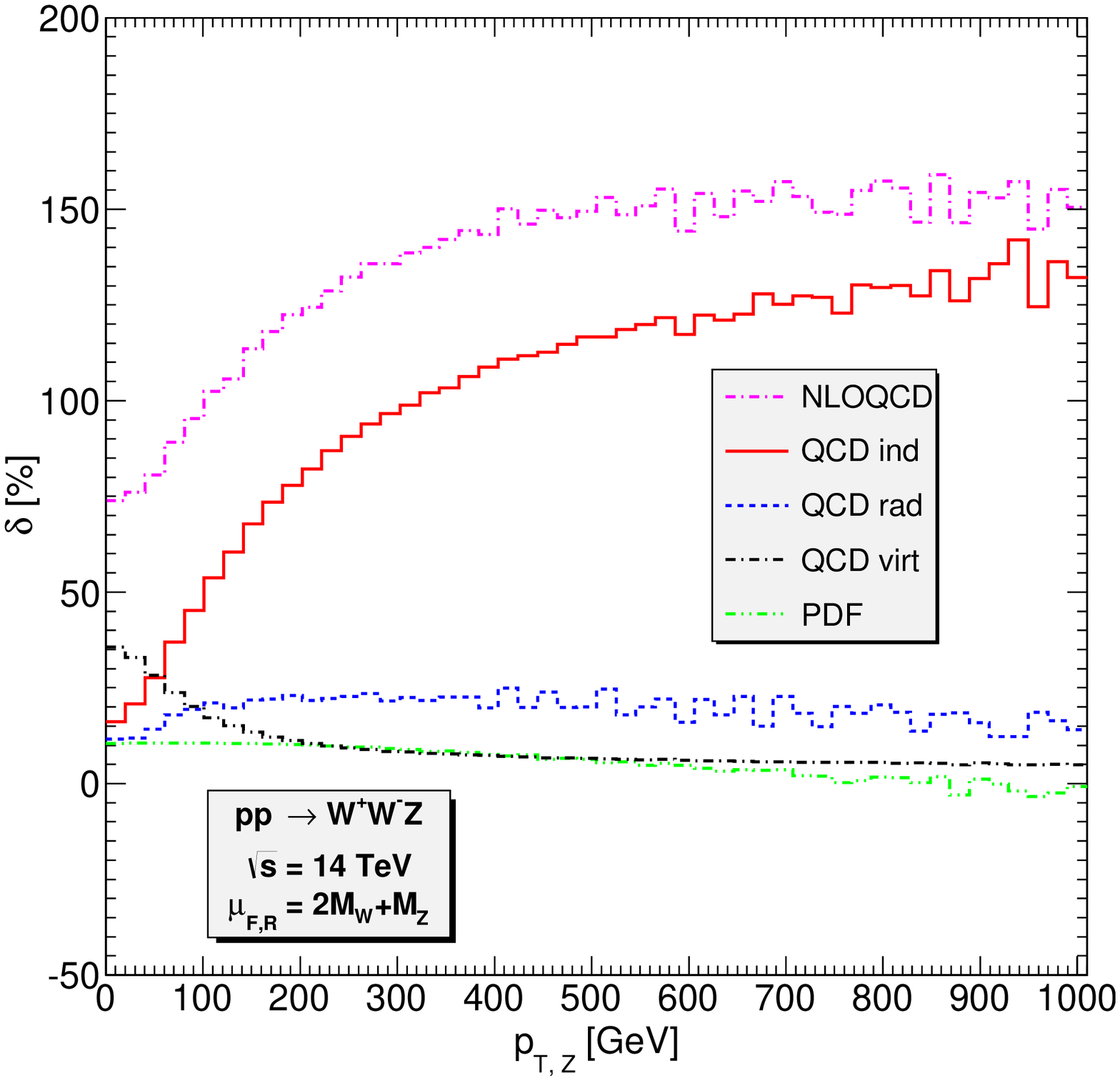}
\includegraphics[width=0.32\textwidth]{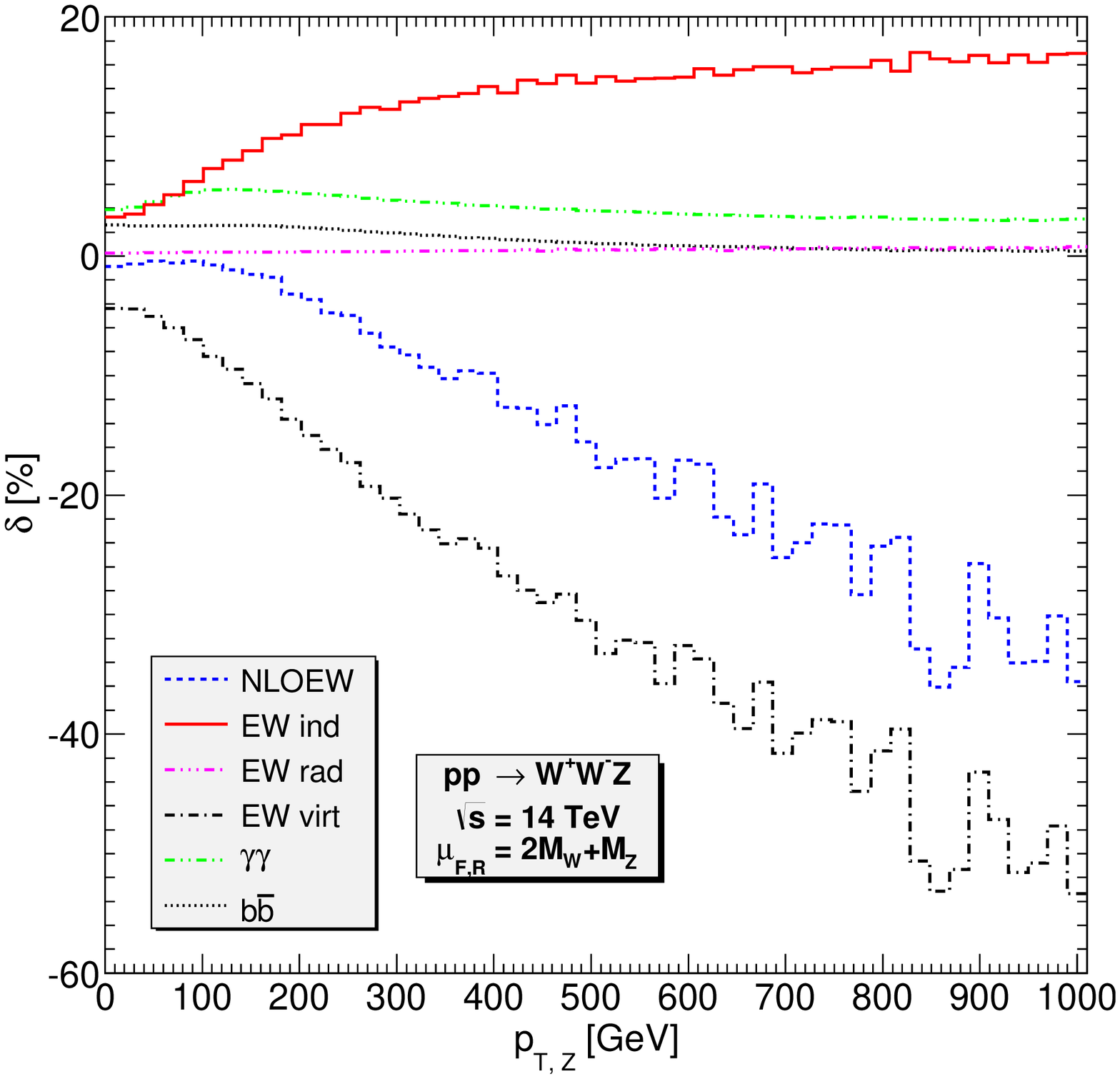}
\caption{$Z$ transverse momentum distribution of $pp \to W^+W^-Z$ cross section (left),
of the NLO QCD corrections (middle) and of the NLO EW corrections (right).
}
\label{qqWWZ_pt}
\end{center}
\end{figure}
%%%
We present the differential cross sections for the LO contribution as well as the
$b\bar b$, $\gamma\gamma$, NLO QCD and NLO EW corrections. The
relative corrections compared to the LO distributions are also shown.
Furthermore, the various QCD and EW subcorrections defined in \sect{total_Xsection} are displayed.

The $Z$ transverse momentum distribution is shown in \fig{qqWWZ_pt}.
From left to right we find the differential cross sections, the NLO QCD and
NLO EW corrections.
The differential cross sections show a maximum at about $p_{T}=50\gev$ and decrease rapidly with $p_T$.
The $b\bar b$ and $\gamma\gamma$ contributions are about 1 to 2 orders of magnitude smaller
than the $q\bar q$ contribution in the whole $p_T$ range.
The NLO QCD correction ($b\bar b$ channel excluded) increases rapidly at low $p_T$ range
and is nearly constant for $p_{T}>400\gev$. The dominant contribution comes from the
gluon-induced subprocesses.
The remaining contributions are less than $30\%$.
The reason for this large gluon-induced correction is that this is a new process with large gluon PDF
opening up at NLO. At large $p_T$, the dominant contribution comes from the mechanism where first
the reaction $ug \to Z u$ with a hard $Z$ balanced by a hard quark occurs. Then, on top of this,
two soft gauge bosons $W^+$ and $W^-$ are radiated. These soft boson radiations introduce
two double logarithms $\alpha^2\log^4(p_{T,Z}^2/M_W^2)$. At LO, the hard $Z$ recoils against
one $W$, hence there is only one double logarithm $\alpha\log^2(p_{T,Z}^2/M_W^2)$ from the soft radiation
of the other $W$. This phenomenon is also observed in $pp \to VV$ with $V=W^\pm, Z$~\cite{Baglio:2013toa}.
While the gluon-induced correction can reach $900\%$ for the $W^-Z$ channel at $p_T = 700\gev$~\cite{Baglio:2013toa},
we get here about $120\%$ for $W^+W^-Z$ production,
which is comparable to the correction in the $ZZ$ case~\cite{Baglio:2013toa}.
For the $W^\pm$ transverse momentum distributions, the correction is smaller.
Moreover, we observe that
the virtual correction rises up in the limit $p_T \to 0$.

\begin{figure}[h]
 \begin{center}
\includegraphics[width=0.32\textwidth]{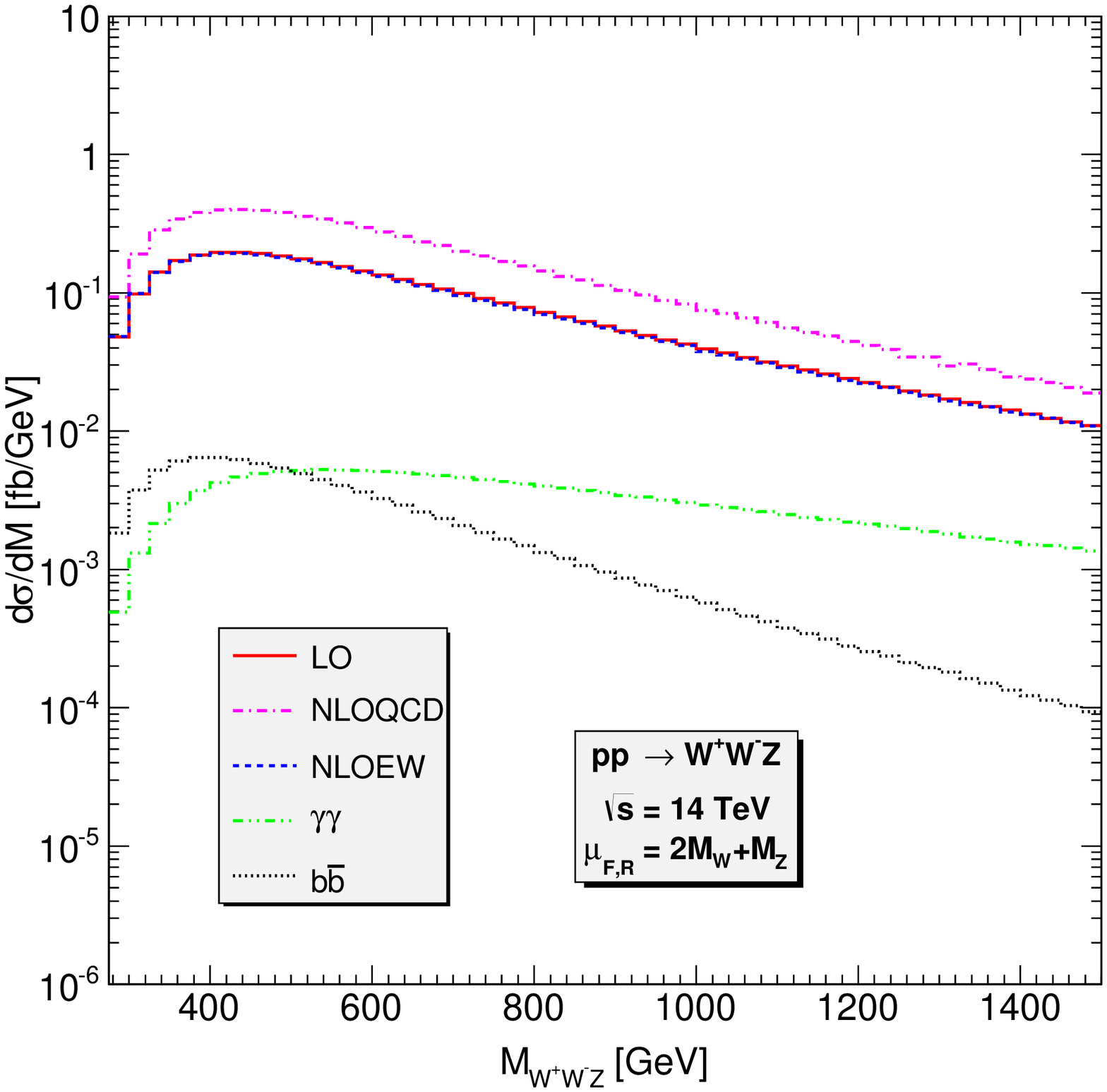}
\includegraphics[width=0.32\textwidth]{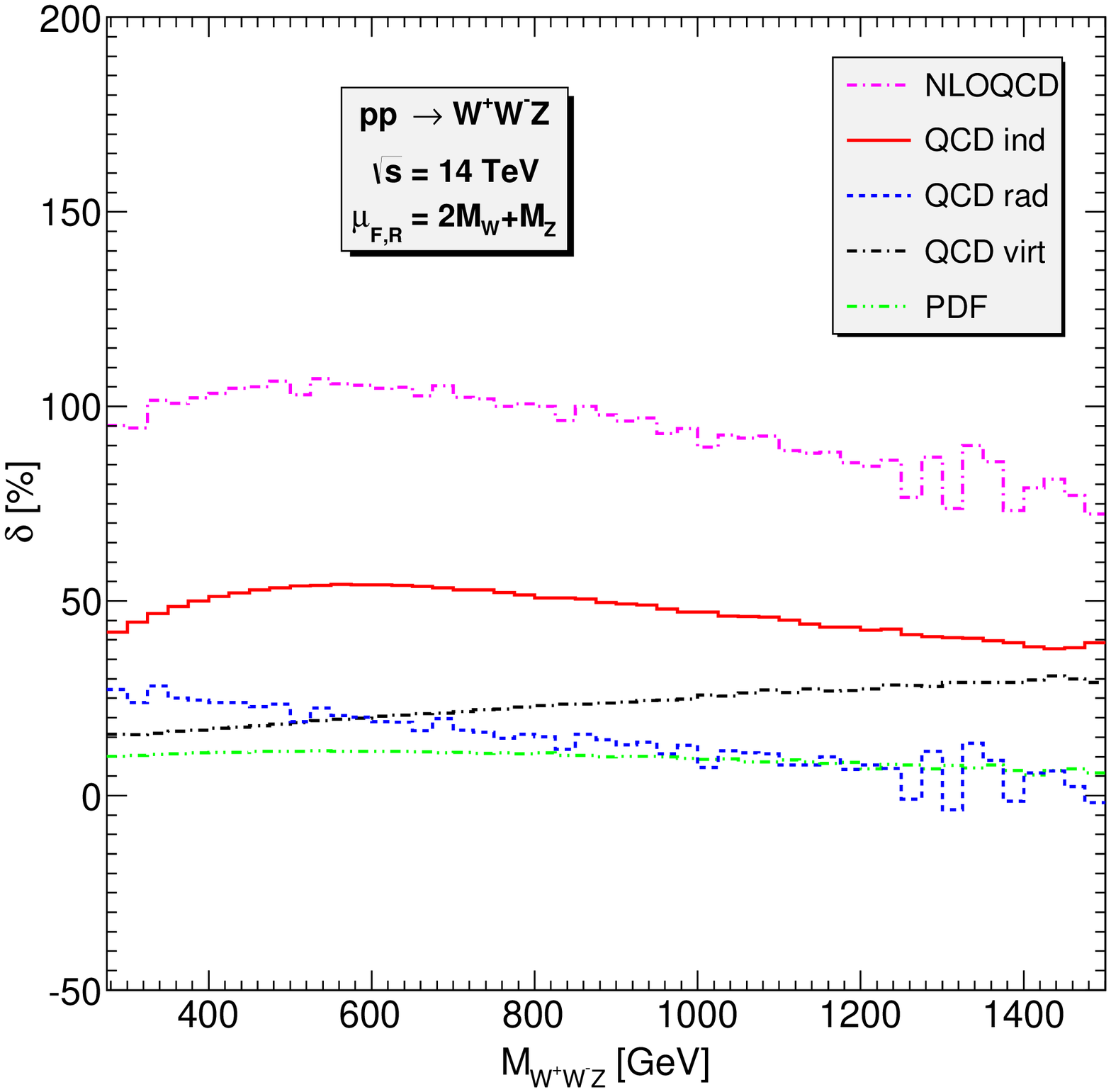}
\includegraphics[width=0.32\textwidth]{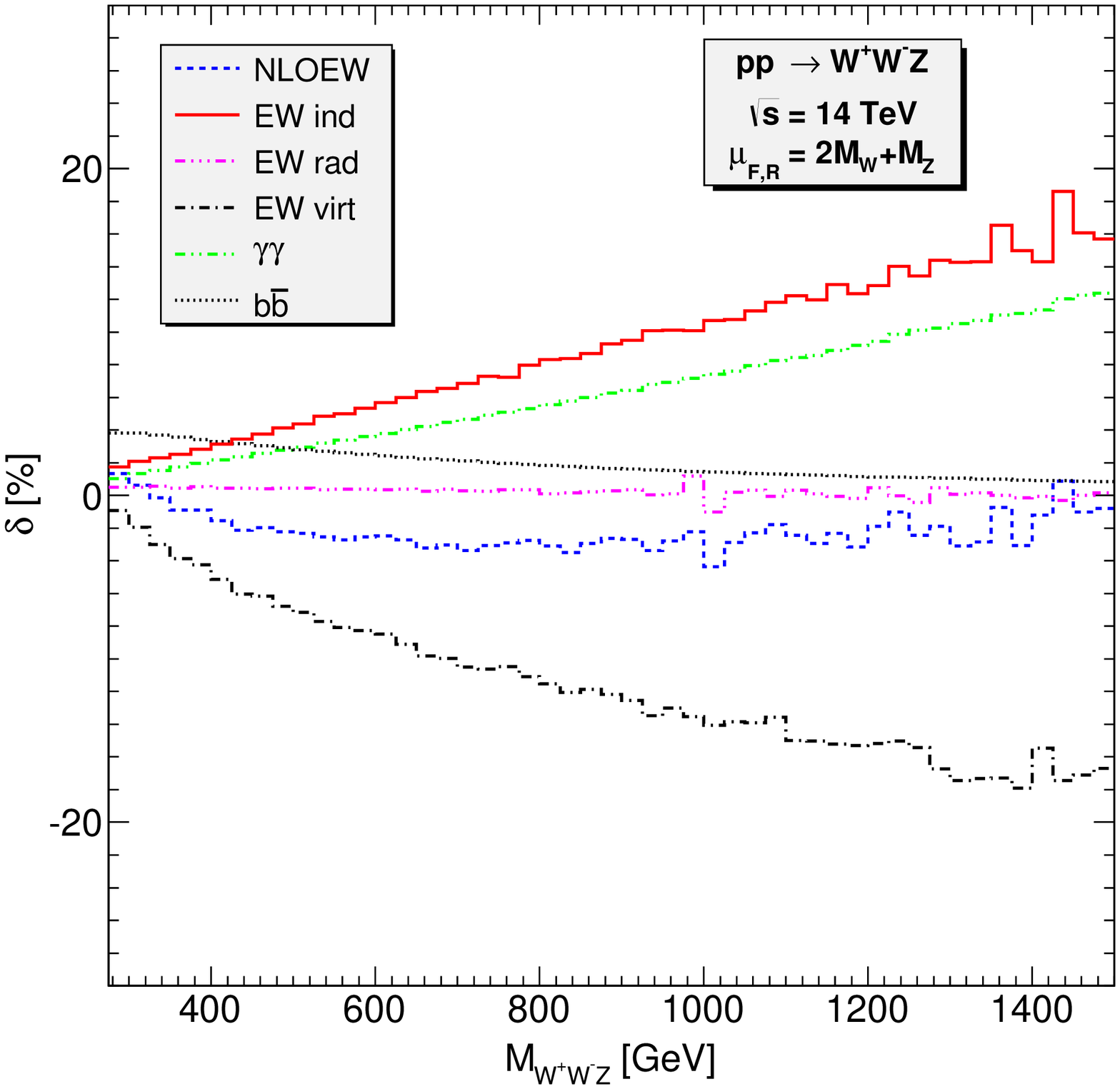}
\caption{Same as \fig{qqWWZ_pt} but for the invariant mass of the $WWZ$ system.}
\label{qqWWZ_M}
\end{center}
\end{figure}
%%%%%
For the NLO EW corrections, the virtual part is negative in the whole $p_T$ range and behaves like $\alpha\log^2(M_V^2/p_T^2)$,
reaching about $-50\%$ at $p_T = 1\tev$.
This is the well-known Sudakov double logarithm arising from the exchange of a virtual massive
gauge boson in the loops.
For the photon-induced correction, the above picture of the gluon-induced correction holds.
There are, however, some important differences. Naively, one would expect that this correction
must be very small because of the EW coupling and small photon PDF,
as it is the case for the $pp \to ZZ$ process~\cite{Baglio:2013toa}. But, as in the case of photon-induced corrections
to $W^\pm Z$ and $W^+ W^-$ production~\cite{Baglio:2013toa}, there is a new enhancement mechanism in the
hard $2 \to 2$ amplitude due to the $t$-channel exchange of a $W$ gauge boson as shown in the first diagram in
\fig{photonic_diag}(b). The hard processes are $qg \to qZ$ for the gluon-induced case, while it
can be $u W^- \to d Z$ for the photon-induced channels. By a simple dimensional analysis, we get
at partonic level and from the $t$-channel diagrams
$|\mathcal{A}_{u W^- \to d Z}|^2/|\mathcal{A}_{u g \to u Z}|^2 \propto E_u^2/q^2$
with $q^2 \approx -2E_u^2(1-\cos\theta)$ being the momentum-transfer square. This enhancement factor
for moderate $q^2$ and some possible additional enhancement from the couplings can lead to a significant
enhancement to compensate for the smallness of the photon PDF. At the end we observe nearly $+20\%$ photon-induced
correction at $p_{T,Z}=1\tev$, canceling part of the Sudakov virtual correction.

In \fig{qqWWZ_M}, we present the invariant mass distribution of the $W^+W^-Z$ system.
For QCD corrections, all contributions are positive and the maximal total correction is
slightly above $100\%$ at $M_{WWZ} = 500\gev$.
Turning to the EW correction plot, we see that the $b\bar{b}$ contribution is important at low
energy while the $\gamma\gamma$ channel is very important at large invariant mass.
The full NLO EW correction ($\gamma\gamma$ and $b\bar{b}$ both excluded) is very small
(less than $4\%$) in the whole range. This is due to the cancellation between the photon-induced
and virtual corrections as shown in the plot.
The $\ga\ga$ correction is larger than the full EW one at large invariant mass.

\begin{figure}[h]
 \begin{center}
\includegraphics[width=0.32\textwidth]{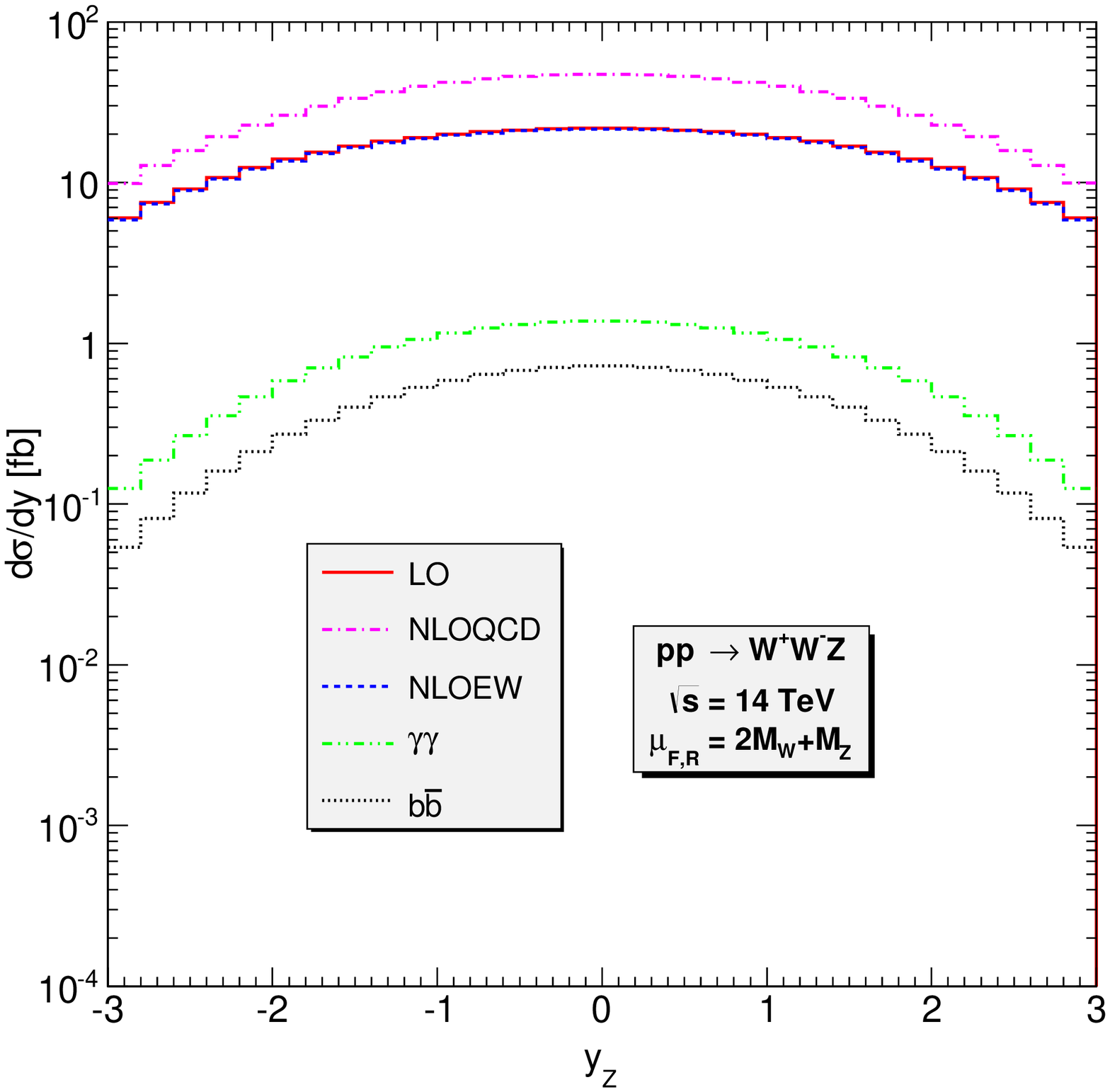}
\includegraphics[width=0.32\textwidth]{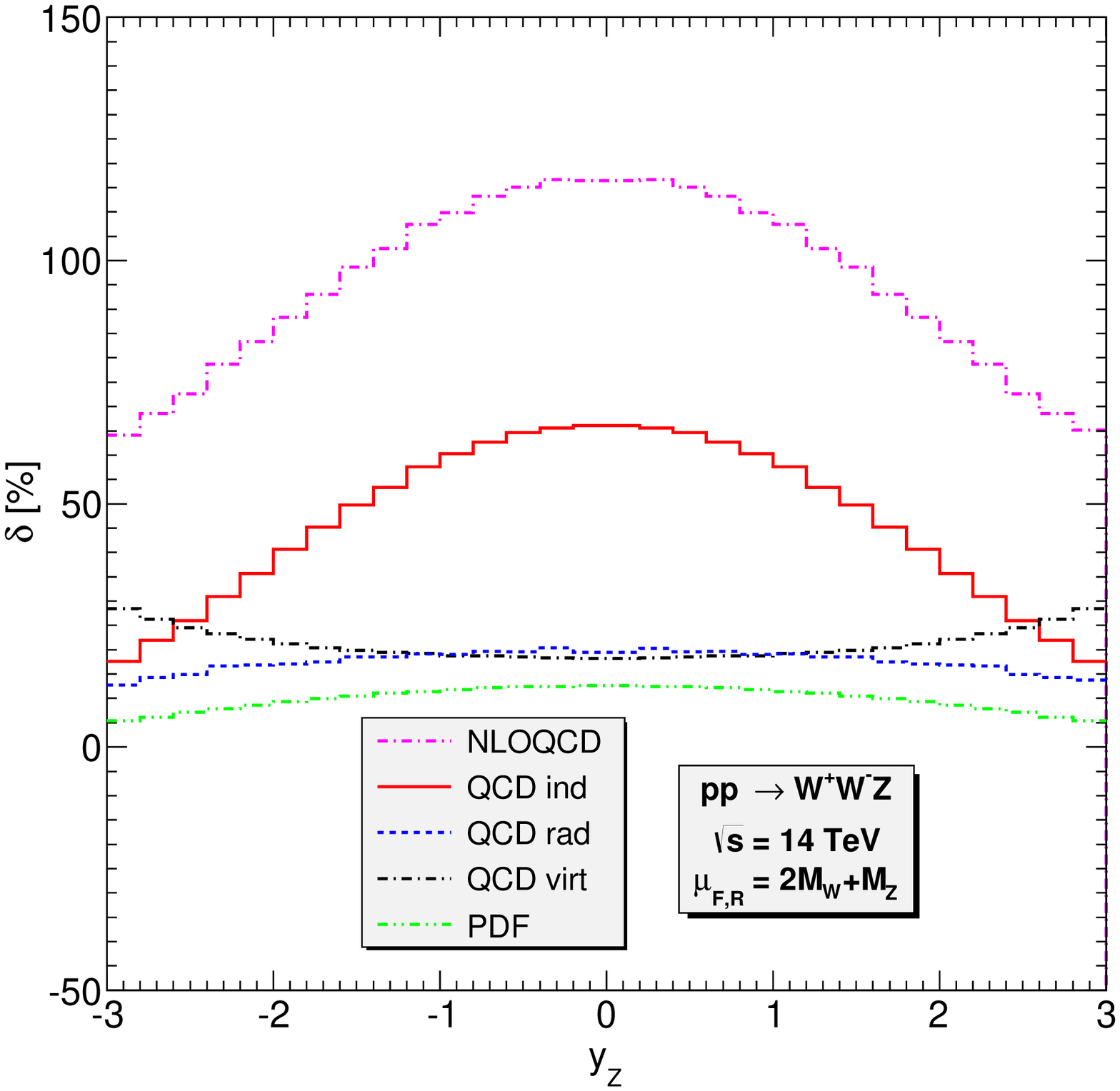}
\includegraphics[width=0.32\textwidth]{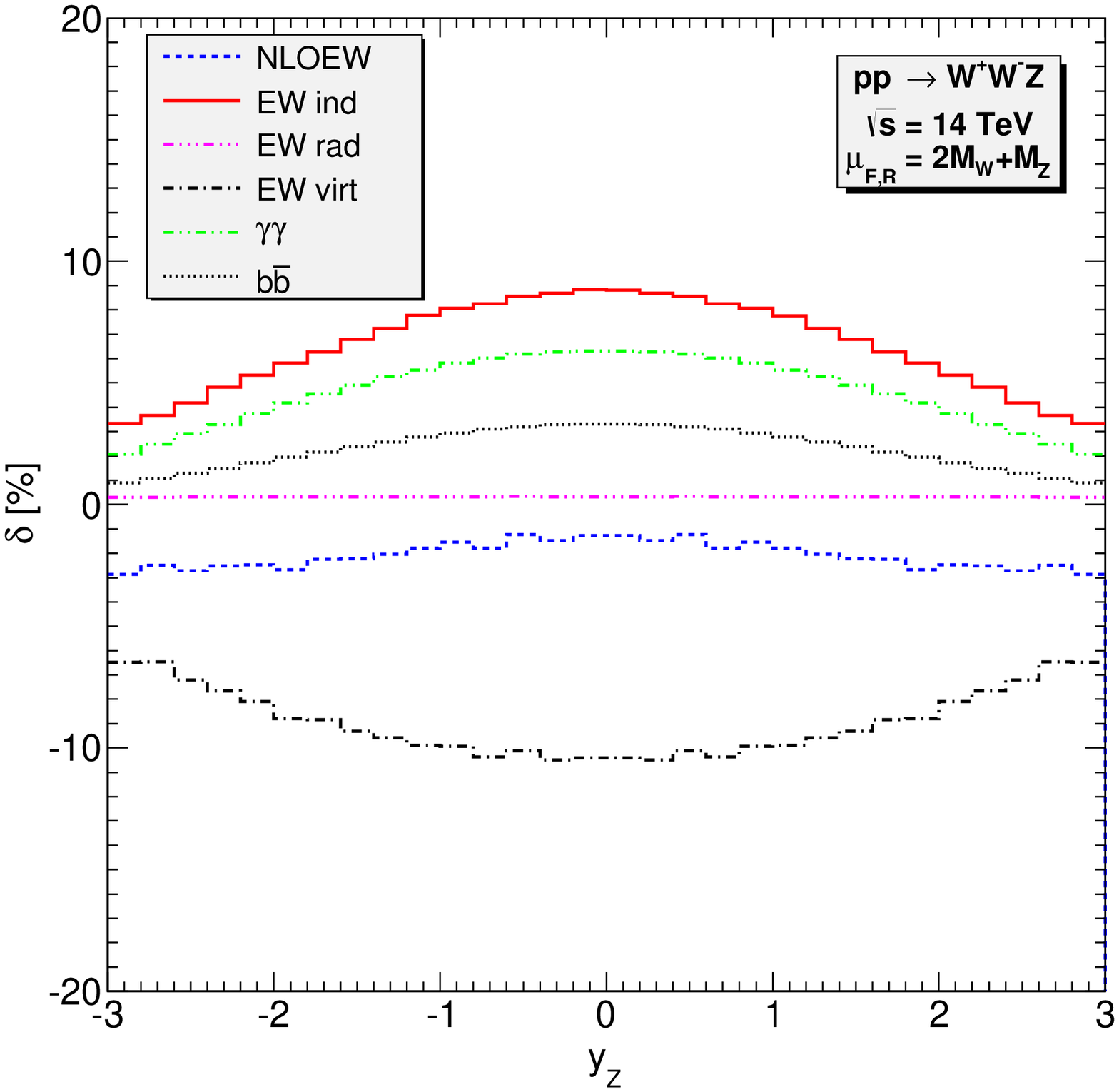}
%%%
\includegraphics[width=0.32\textwidth]{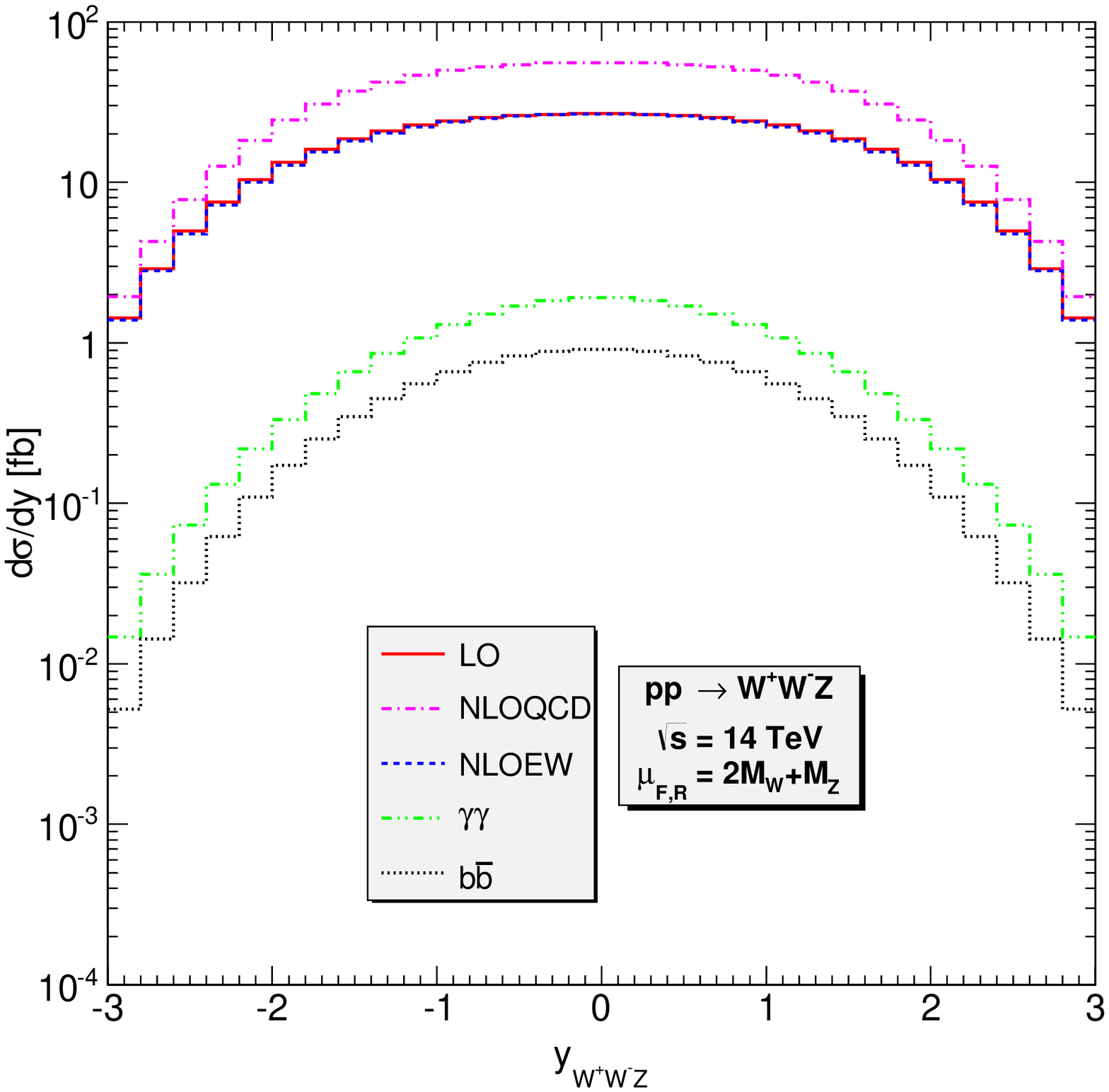}
\includegraphics[width=0.32\textwidth]{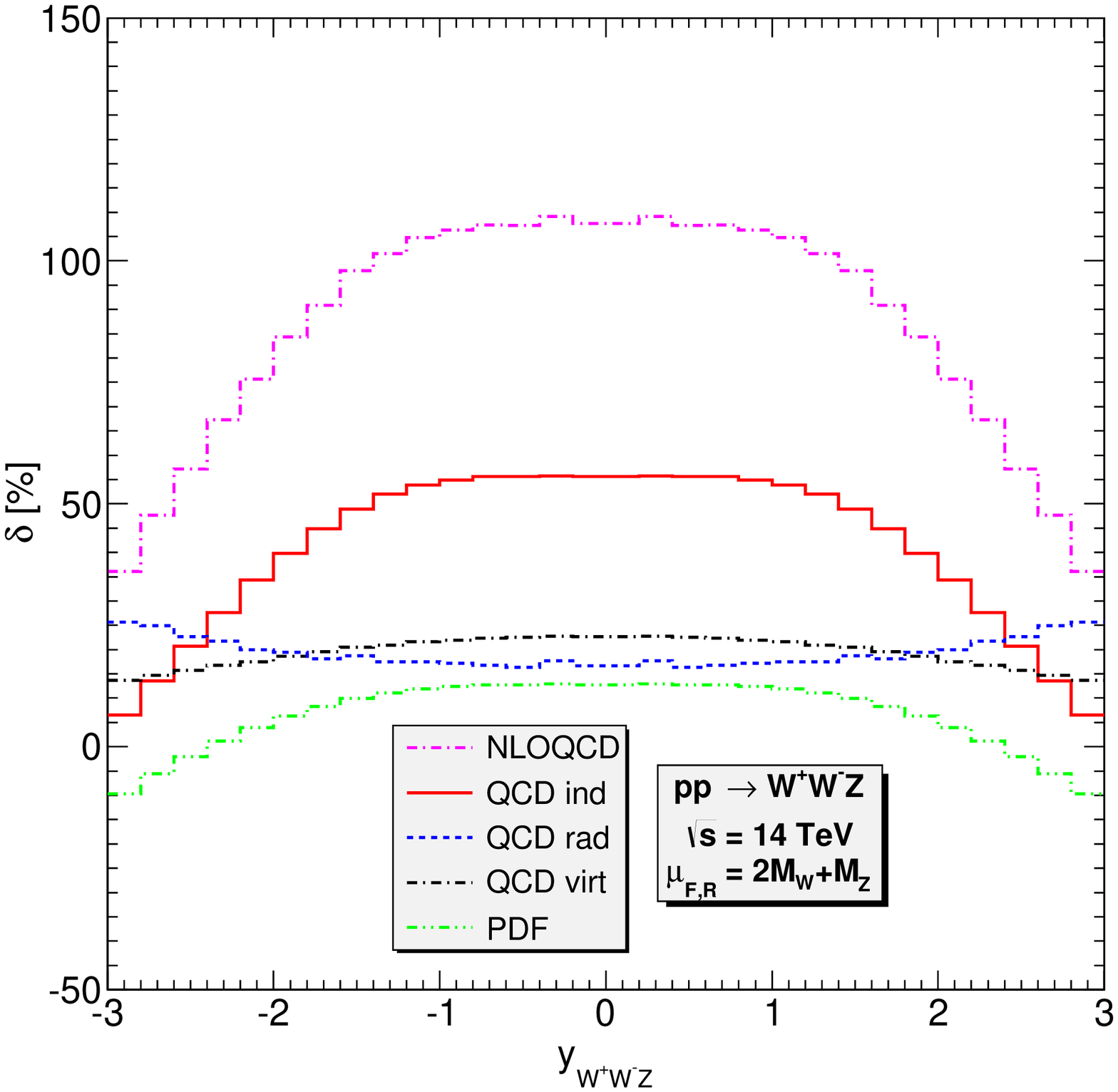}
\includegraphics[width=0.32\textwidth]{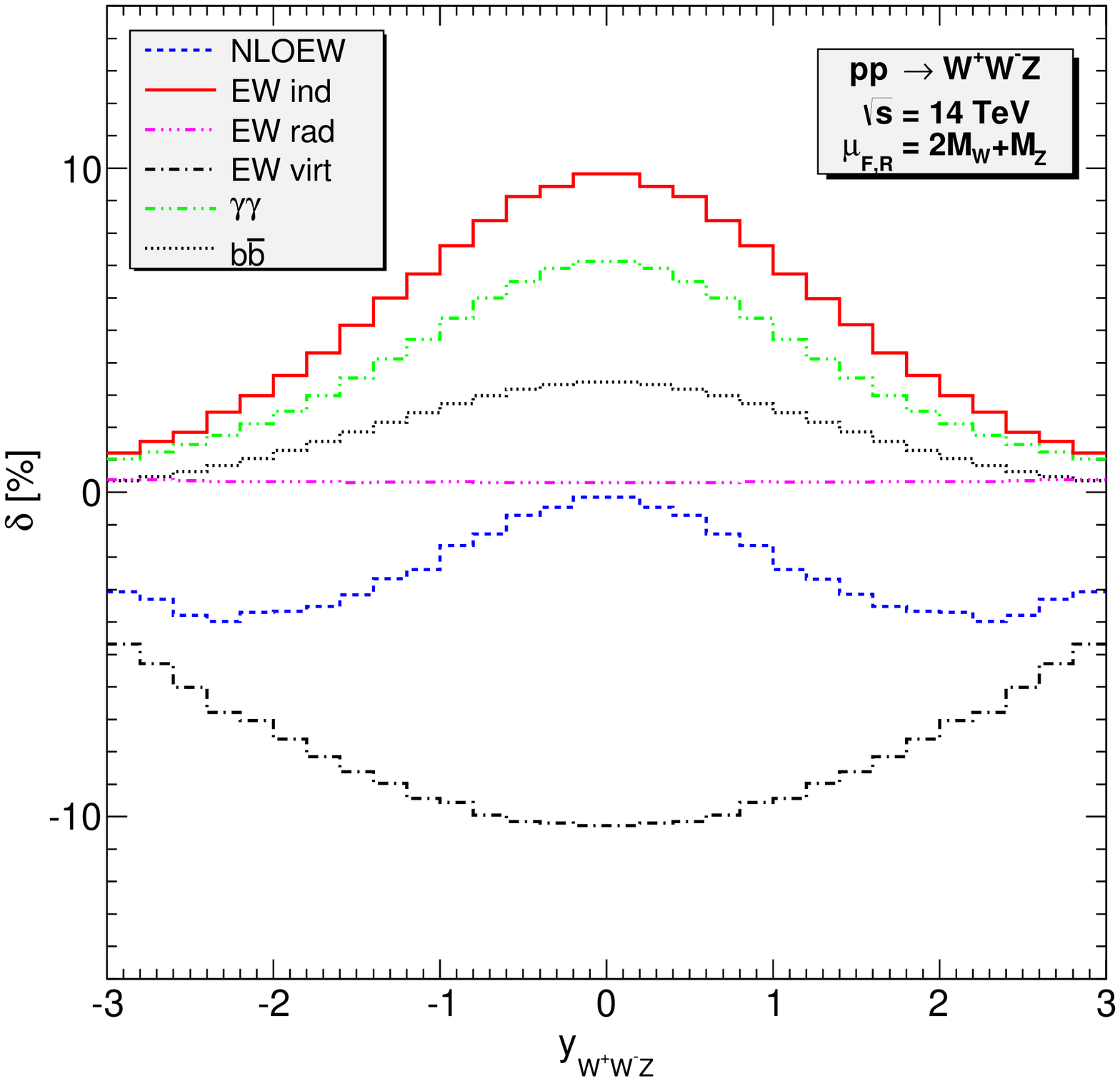}
\caption{Same as \fig{qqWWZ_pt} but for the rapidity of the $Z$ (first row) and of the $WWZ$ system (second row).}
\label{qqWWZ_y}
\end{center}
\end{figure}
%%%%%%
We next display in \fig{qqWWZ_y} the rapidity distribution of the $Z$ boson in the first row
and of the $W^+W^-Z$ system in the second row. One can see that both the $Z$ boson and the
$W^+W^-Z$ system are centrally produced. In both cases, the QCD correction is dominated by
the gluon-induced contribution and maximal in the central region. For the EW correction plot,
we again see the importance of the $\gamma\gamma$ channel and the cancellation between the photon-induced
and virtual corrections. In both cases, the full EW correction is negative and its magnitude is always
less than $5\%$.

\begin{figure}[h]
 \begin{center}
\includegraphics[width=0.6\textwidth]{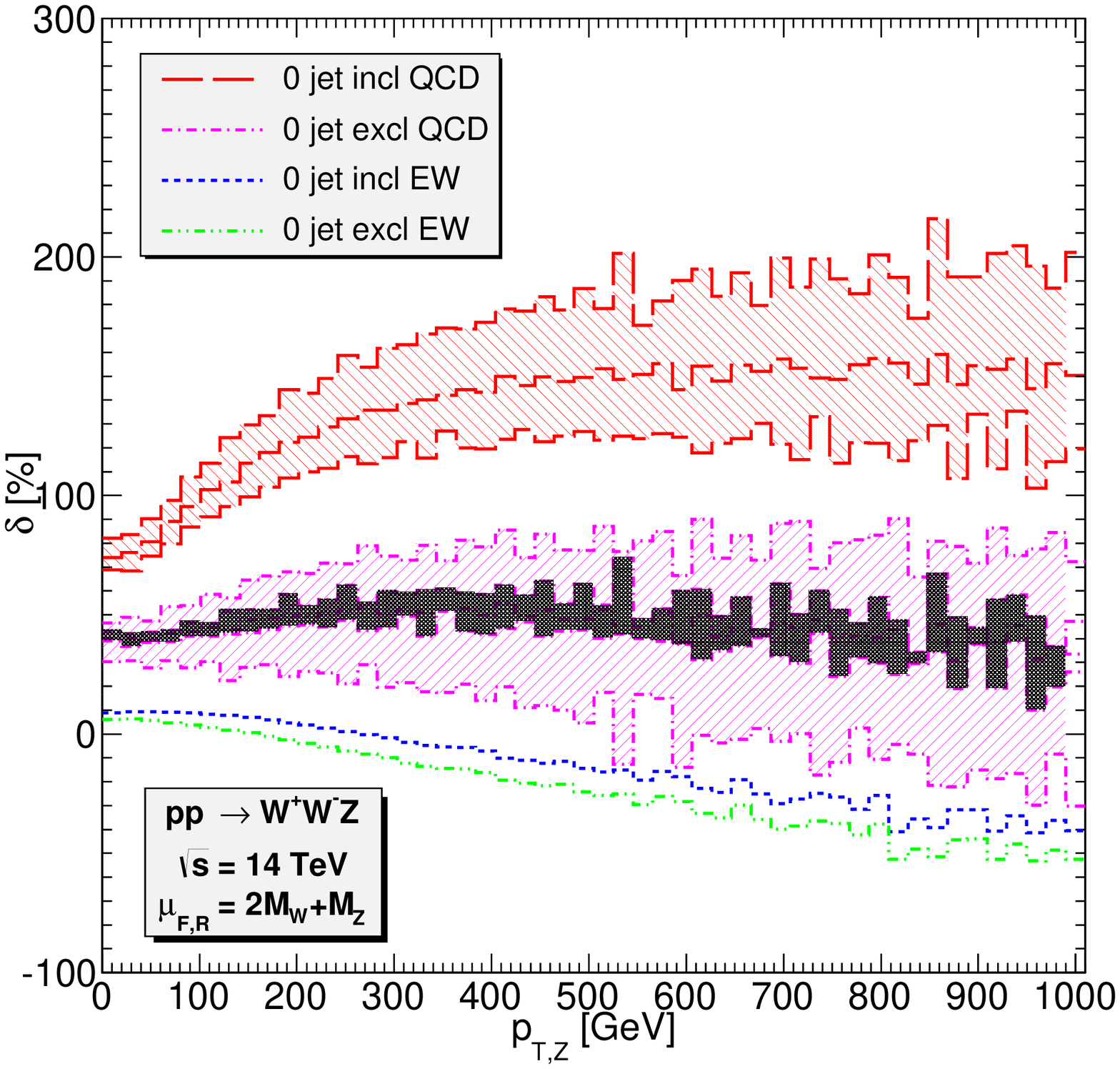}
\caption{NLO QCD and EW corrections to the $Z$ transverse momentum distribution for
inclusive events without jet cuts and also for exclusive events with a dynamic
jet veto defined in the text.
The bands describe $\mu_0/2 \le \mu_F = \mu_R \le 2\mu_0$ with $\mu_0 = 2M_W+M_Z$ variations 
of the NLO QCD corrections. The LO result is calculated at the central scale everywhere. 
The band at the top is for the inclusive distribution.
The two other bands are for the exclusive distribution. Their definitions are given in the text.
}
\label{fig:jetveto_dyn}
\end{center}
\end{figure}
%%%%%%%%%
From the above phase-space dependence study, we see that the NLO QCD correction mainly due to the
$2 \to 4$ gluon-induced channels is very large at high $p_T$. The dominant contribution comes from
the region where the quark transverse momentum is large. It is therefore attractive to think of
imposing a jet veto to reduce this large QCD contribution,
as done for example in Refs.~\cite{Campanario:2008yg,Denner:2009gj}.
One should be very careful in doing so because using a jet veto increases theoretical
uncertainty due to missing large higher-order corrections~\cite{Stewart:2011cf}. In this sense, the cross section
with jet veto is less perturbative than the inclusive cross section. This view is
supported in \fig{fig:jetveto_dyn}. Here we apply a dynamic jet veto: for the exclusive
zero-jet distribution, we veto events with $p_\text{T,jet} > p_\text{veto}$, with
\bea
p_\text{veto} = \fr{1}{2} \text{max}(M_{\text{T},W^+},M_{\text{T},W^-},M_{\text{T},Z}),
\label{dyn_veto}
\eea
where $M_{\text{T},V}=(p_{\text{T},V}^2 + M_{V}^2)^{1/2}$ is the transverse mass.
We have tried a fixed jet veto with $p_\text{veto} = 25\gev$ and
found that it over removes the NLO QCD correction,
leading to large negative QCD correction at high $p_{T,Z}$. With the dynamic jet veto, we found that
more than half of the QCD correction is removed. However, the uncertainty band on the
exclusive zero-jet distribution is larger than the band on the inclusive zero-jet distribution.
The reason is the following. We have
\bea
d\sigma_\text{0j,inc} = d\sigma_\text{0j,exc} + d\sigma_\text{1j,inc}.
\label{inc_exc_eq}
\eea
The inclusive zero-jet distribution $d\sigma_\text{0j,inc}$ is independent of $p_\text{veto}$, while
both the exclusive zero-jet $d\sigma_\text{0j,exc}$ and inclusive one-jet $d\sigma_\text{1j,inc}$
distributions depend on $\log(p_\text{veto}/p_{T,Z})$.
The two terms in the right-hand side of \eq{inc_exc_eq} are therefore not independent.
Thus, as argued in Ref.~\cite{Stewart:2011cf}, it is suitable to consider $d\sigma_\text{0j,inc}$ and $d\sigma_\text{1j,inc}$ as
independent observables and calculate $d\sigma_\text{0j,exc}$ from them. This means that the
scale uncertainty of the exclusive zero-jet distribution is calculated as
\bea
\Delta^2_\text{0j,exc} = \Delta^2_\text{0j,inc} + \Delta^2_\text{1j,inc}.
\label{inc_exc_Delta}
\eea
This explains the large uncertainty band (in pink) of the exclusive distribution.
In passing, we also show the naive uncertainty band (the smallest band in black) calculated as
$\Delta_\text{0j,exc} = \Delta_\text{0j,inc} - \Delta_\text{1j,inc}$ assuming
that the two inclusive observables are anti-correlated.

In \fig{fig:jetveto_dyn}, we also show the effect of the dynamic jet veto on the EW correction.
Here, for the photon-radiated contribution, the photon is treated as a jet. We observe a small effect.
For the EW correction, there is no uncertainty band because the scale dependence is of QCD origin as
pointed out in \sect{total_Xsection}.

%%%%%%%%%%%%%%%%%%%%%%%%%%%%%%
\section{Conclusions}        %
\label{sect-conclusions}     %
%%%%%%%%%%%%%%%%%%%%%%%%%%%%%%
In this paper, we have presented
the first calculation of the  NLO EW correction in combination with the NLO QCD correction
to the $W^+W^-Z$ production at the LHC with $14\tev$ center-of-mass energy.
This provides the most up-to-date prediction of the total and differential cross sections.
The NLO QCD correction is large, about $+100\%$ for the total cross section.
For EW correction, not only the photon-radiated but also the photon-induced contributions
are taken into account. The latter turns out to be important and cancel
part of the large Sudakov virtual correction. This leads to very small EW correction, about $-2\%$,
for the total cross section.
This cancellation happens, to varying extent, also in the transverse momentum, invariant mass and rapidity distributions.

We have also discussed the use of a jet veto to reduce the large QCD correction. We found that
using a dynamic jet veto is good in the sense that it allows the jet to be away from the non-perturbative
regime and removes significantly the QCD correction. On other hand, it increases QCD uncertainty due to
missing large higher-order corrections.

\bigskip

\noindent{\bf Acknowledgments:}
We thank Dieter Zeppenfeld for fruitful discussions.
This work is supported by the Deutsche
Forschungsgemeinschaft via the 
Sonderforschungsbereich/ Transregio 
SFB/TR-9 Computational Particle Physics.

\appendix

\section{Results at one phase-space point \label{appendixA}}
In this appendix we provide results at a random phase-space point to facilitate comparisons
with our results, in particular for those trying to develop automated tools. The phase-space point
for the process $\bar{q} q \to W^+ W^- Z$ is given in \tab{table_PSP_2to3}.
\begin{table}[h]
 \begin{footnotesize}
 \bc
 \caption{\label{table_PSP_2to3}{A random phase-space point for $\bar{q} q \to W^+ W^- Z$ subprocesses.}}
%\vspace*{0.5cm}
\begin{tabular}{l | r@{.}l r@{.}l r@{.}l r@{.}l}
% \hline
& \multicolumn{2}{c}{ $E$}
& \multicolumn{2}{c}{ $p_x$}
& \multicolumn{2}{c}{ $p_y$}
& \multicolumn{2}{c}{ $p_z$}
\\
\hline
$\bar{q}$  & 234&035328935400 & 0&0  & 0&0  &  234&035328935400  \\
$q$  & 234&035328935400 & 0&0  & 0&0  &  -234&035328935400  \\
$W^+$  & 204&344376484520 & -120&509782379302 & 28&2759628195356 & 141&324938540120 \\
$W^-$  & 133&625238535211 & 87&1775591913742 & -28&2759628195356 & -54&7220179512301 \\
$Z$  & 130&101042851068 & 33&3322231879280 & 0&0 & -86&6029205888900 \\
\hline
\end{tabular}\ec
 \end{footnotesize}
\end{table}
%%%
In the following we provide the squared amplitude with the averaged factor
over helicities and colors. We also set $\alpha = \alpha_s = 1$ for simplicity.
At tree level, we have
\begin{align}
\overline{|\mathcal{A}_\text{LO}^{\bar{u}u}|}^2 &= 0.961753014217244,\crn
\overline{|\mathcal{A}_\text{LO}^{\bar{d}d}|}^2 &= 12.3829496659527.
\end{align}
The interference amplitudes 
$2\text{Re}(\mathcal{A}_\text{NLO}\mathcal{A}^{*}_\text{LO})$, 
for the virtual QCD corrections defined in \eq{xsection_virt_qcd}, 
are given in \tab{table_PSP_QCD_U} and \tab{table_PSP_QCD_D}. Here we use the following
convention for one-loop integrals, with $D=4-2\eps$,
\bea
T_0 = \fr{\mu^{2\eps}\Gamma(1-\eps)}{i\pi^{2-\eps}}\int d^D q \fr{1}{(q^2 - m_1^2 + i0)\cdots}.
\eea
This amounts to dropping a factor ${(4\pi)^\epsilon}/{\Gamma(1-\epsilon)}$ 
both in the virtual corrections and the I-operator.
%%%
\begin{table}[h]
 \begin{footnotesize}
 \bc
\caption{\label{table_PSP_QCD_U}{QCD interference amplitudes $2\text{Re}(\mathcal{A}_\text{NLO}\mathcal{A}^{*}_\text{LO})$
for terms in \eq{xsection_virt_qcd} for $\bar{u} u \to W^+ W^- Z$ subprocess.}}
%\vspace*{0.5cm}
\begin{tabular}{l | r@{.}l r@{.}l r@{.}l}
% \hline
& \multicolumn{2}{c}{ $1/\eps^2$}
& \multicolumn{2}{c}{ $1/\eps$}
& \multicolumn{2}{c}{ finite}
\\
\hline
QCD-I  & 0&408180656656545 & 0&106650712644880  & -0&418657743041666 \\
QCD-loop  & -0&408180656656539 & -0&106650712644797 & 1&63036547637921  \\
QCD-virt  &  5&307828480419084$\times 10^{-15}$ & 8&387679573995589$\times 10^{-14}$ & 1&21170773333755 \\
\hline
\end{tabular}\ec
 \end{footnotesize}
\end{table}
%%%
\begin{table}[h]
 \begin{footnotesize}
 \bc
 \caption{\label{table_PSP_QCD_D}{QCD interference amplitudes $2\text{Re}(\mathcal{A}_\text{NLO}\mathcal{A}^{*}_\text{LO})$
for terms in \eq{xsection_virt_qcd}
for $\bar{d} d \to W^+ W^- Z$ subprocess.}}
%\vspace*{0.5cm}
\begin{tabular}{l | r@{.}l r@{.}l r@{.}l}
% \hline
& \multicolumn{2}{c}{ $1/\eps^2$}
& \multicolumn{2}{c}{ $1/\eps$}
& \multicolumn{2}{c}{ finite}
\\
\hline
QCD-I  &  5&25548706505201 & 1&37317002078168  & -5&39038368760993 \\
QCD-loop  & -5&25548706505202 & -1&37317002078186 & 19&8522399631644  \\
QCD-virt  & -3&145379840248346$\times 10^{-15}$ & -1&813835707876546$\times 10^{-13}$ & 14&4618562755545 \\
\hline
\end{tabular}\ec
 \end{footnotesize}
\end{table}
%%%
Moreover, the dimensional regularization method~\cite{'tHooft:1972fi, Hahn:1998yk}
with $\mu_{F} = \mu_{R} = 2M_W + M_Z$ is used.
For the dimensional reduction scheme, the $\text{I}$-operator and loop amplitudes are different, but
their sum must be the same~\cite{Catani:1996pk}. The finite part of the
virtual QCD correction is independent of $\mu_{F}$ and $\mu_{R}$.

For EW corrections, we use mass regularization and the results are given in
\tab{table_PSP_EW_U_D}. Note that, as written in \sect{cal_ew}, the $\text{I}$-operator
contribution is now defined as the endpoint contribution in \bib{Dittmaier:1999mb}.
The light fermion mass regulator is $m_f = 10^{-4}\gev$ (with $f\neq t$).
We have checked that the virtual EW correction, \ie\ the sum of the $\text{I}$-operator
and loop contributions, is UV and IR finite as well as independent of $m_f$.
If we change to $m_f = 10^{-3}(10^{-5})\gev$ then we obtain $8(10)$ digit agreement using
double precision.
%%%
\begin{table}[h]
 \begin{footnotesize}
 \bc
 \caption{\label{table_PSP_EW_U_D}{Interference amplitudes $2\text{Re}(\mathcal{A}_\text{NLO}\mathcal{A}^{*}_\text{LO})$
for EW corrections as defined in the text.}}
%\vspace*{0.5cm}
\begin{tabular}{l | r@{.}l r@{.}l}
% \hline
& \multicolumn{2}{c}{ $\bar{u} u \to W^+ W^- Z$}
& \multicolumn{2}{c}{ $\bar{d} d \to W^+ W^- Z$}
\\
\hline
EW-I  & -8&09003628219715 & -34&6814203416028  \\
EW-loop  & -10&5259914893826 & -70&1705883597006  \\
EW-virt  & -18&6160277715797 & -104&852008701303 \\
\hline
\end{tabular}\ec
 \end{footnotesize}
\end{table}
%%%

%\bibliographystyle{h-physrev}
%\bibliography{main}

\end{document}